\documentclass[10pt,a4paper]{article}

\font \eightrm=cmr8

\usepackage{amssymb}
\newtheorem{thm}{Theorem}
\newtheorem{defn}{Definition}
\newtheorem{cor}{Corollary}
\newtheorem{lem}{Lemma}

\newtheorem{prop}{Proposition}

\newtheorem{rk}{Remark}
\usepackage{amssymb}
\newcommand\cutoffsum{-\hskip -4mm\sum}

\def\dbar{d{\hskip-1pt\bar{}}\hskip1pt}

\def \C{{\! \rm \
I \!\!\!C}}
\def \R {{\! \rm \ I \!R}}
\def \endsquare{ $\sqcup \!\!\!\! \sqcap$ }
\def \N {{\! \rm \ I \!N}}

\def \Z {{\! \rm Z\! \!Z}}
 \def\cutoffint{-\hskip -10pt\int}
\def\altcutoffint{=\hskip -11pt\int}

\def\otherterm#1{{\it#1}}

\def \Ci {{C^\infty}}

\def \fp{{\rm f.p.}}

\begin{document}

\title{\bf Shuffle relations for regularised integrals of symbols}

\author{  Dominique MANCHON, Sylvie
PAYCHA }

\maketitle
\section*{Abstract}
 We prove  shuffle relations which  relate a  product of regularised  integrals of classical symbols $ \int^{reg} \sigma_i\, d\xi_i, i=1, \cdots, k$   to  regularised nested iterated
 integrals: 
 $$\prod_{i=1}^k  \int^{reg} \sigma_i\, d\xi_i=\sum_{\tau \in \Sigma_k}
 \int^{reg} d\xi_1 \int_{\vert \xi_{2}\vert \leq \vert  \xi_1\vert} d\xi_{2}
 \cdots \int_{\vert \xi_L\vert \leq \vert \xi_{k-1}\vert} d\xi_k\,
 \otimes_{i=1}^k \sigma_{\tau(i)}$$  where $\Sigma_k$ is the group of
 permutations over $k$ elements. We show that these shuffle relations   hold
 if  all the  symbols $\sigma_i$ have  vanishing residue;  this is true of
 non integer order symbols on which the regularised integrals have all the
 expected  properties  such as Stokes' property \cite{MMP}. In general  the
 shuffle relations  hold up to finite parts of corrective terms  arising from  a renormalisation on  tensor products of classical symbols, a procedure  adapted from renormalisation procedures on Feynman diagrams  familiar to physicists.\\
We relate the shuffle relations for regularised integrals of symbols with
shuffle relations for multiple zeta functions adapting  the above constructions to
the case of a symbol on the unit circle.    
\section*{Acknowledgements}
The first author would like to thank  K. Ebrahimi-Fard  for stimulating
discussions around renormalisation and both authors are thankful to Li Guo
for interesting discussions around the Rota-Baxter relation, which served as a
motivation for
parts of this work. We also wish to thank D. Kreimer for pointing out to us
some valuable references.  The second author is grateful to the Max Planck
Institute 
where parts of this article were worked out and to D. Zagier with whom some
informal discussions around  zeta functions associated with operators took
place as well as to M. Marcolli for stimulating  discussions on
 renormalisation and G. Racinet for his patient  explanations concerning
 multiple zeta functions.   
\vfill \eject \noindent

\section*{Introduction}
Before  describing the contents of the paper, let us  give  some general motivation. 
Starting from a function $f:\N\to \C$, one can build  functions $P(f): \N\to \C$ and  $\tilde P(f): \N\to \C$:
$$P(f)(n)= \sum_{n>m>0} f(m), \quad \tilde P(f)(n)= \sum_{n\geq m> 0} f(m).$$
The operators $P$ and $\tilde P$  obey  Rota-Baxter relations and define Rota-Baxter type operators of weight $-1$ and $1$ respectively:
$$P(f) \, P(g)= P\left(f\, P(g)\right)+ P\left( g\, P(f)\right) + P(fg)$$
 and 
$$\tilde P(f) \,\tilde  P(g)= \tilde P\left(f\, \tilde P(g)\right)+\tilde P\left( g\,\tilde P(f)\right) - \tilde P(fg).$$
 When applied to $f(n)= n^{-z_1}$, $g(n)=n^{-z_2}$, these relations lead to
 the ``second shuffle relations'' for zeta functions \cite{ENR}:
$$\zeta(z_1)\, \zeta(z_2)= \zeta(z_1, z_2) +\zeta(z_2,z_1) +\zeta(z_1+z_2)$$
where $\zeta(z)= \sum_{n>0} n^{-z}$ and $\zeta(z_1, z_2)= \sum_{n_1>n_2}
n_1^{-z_1} n_2^{-z_2}$. Similarly, 
$$\zeta(z_1)\, \zeta(z_2)= \tilde \zeta(z_1, z_2) +\tilde \zeta(z_2,z_1) -\tilde \zeta(z_1+z_2)$$
where  $\tilde \zeta(z_1, z_2)= \sum_{n_1\geq n_2} n_1^{-z_1} n_2^{-z_2}$.\\
Correspondingly, starting from $f\in L^1(\R, \C)$, one can build $P(f): R\to \C$:
$$P(f)(y)= \int_{y\geq x } f(x)\, dx.$$
Then the classical Rota-Baxter relation (of weight zero)
$$P(f)\, P(g)= P(f\, P(g))+ P(g \, P(f))$$ 
 is an integration by parts in disguise. It leads to  
 to shuffle relations for integrals:
$$\prod_{i=1}^k \int_\R f_i= \sum_{\tau \in \Sigma_k} \int_\R P\left(
  P\left(\cdots P(f_{\tau(k)}) \, f_{\tau(k-1)}\cdots\right)
  f_{\tau(2)}\right)\,f_{\tau(1)}\quad  \forall k\geq 2 $$
under adequate integrability assumptions on the functions $f_i$. \\
Zeta functions generalize to zeta functions associated to  elliptic classical pseudo-differential  operators on a closed manifold $M$  defined by 
 $$\zeta_{A}(z)=\sum_{\lambda_n\in {\rm Spec}(A), \lambda_n\neq 0}\lambda_n^{-z}$$
modulo  some extra under assumptions  on the leading symbol of  the operator
$A$  to ensure the existence of its complex power $A^{-z}$. If $\sigma_A(z)$
denotes the symbol of this complex power   then   provided the order of $A$ is
positive, for Re$(z)$ large
enough, $\zeta_A$  is actually an integral of the symbol on the cotangent bundle $T^*M$:
$$\zeta_A(z)= \int_M \, dx \int_{T_x^*M} {\rm tr}_x(\sigma_A(z))(x, \xi) \,  \dbar\xi$$
with $\dbar \xi:= \frac{d\xi}{(2\pi)^n}$, $n$ being the dimension of $M$. It extends to a meromorphic function on the whole plane replacing the ordinary integral by a cut-off integral $\cutoffint_{T^*M}$. \\ \\
The main purpose of this paper is to establish shuffle relations for cut-off
integrals of classical symbols $\sigma_i\in CS^{\alpha_i}(U_i)$ (see notations
in the Preliminaries):
$$\prod_{i=1}^k \cutoffint  \sigma_i = \sum_{\tau \in \Sigma_k} \cutoffint
P\left(\cdots P\left( P(\sigma_{\tau(k)}) \, \sigma_{\tau(k-1)}\right)\cdots
  \sigma_{\tau(2)}\right) \sigma_{\tau(1)}\quad \forall k\geq 2$$
and other regularised integrals built from cut-off integrals. We give 
sufficient assumptions on the symbols for such shuffle relations to hold,
conditions which we shall specify below, once we have introduced the necessary
technical tools. It turns out that on the class of  non integer order classical
symbols, on which these regularised integrals have the expected properties
such as Stokes' property, translation invariance...(see \cite{MMP}), these shuffle
relations hold. 
Otherwise a renormalisation procedure is needed to take care of obstructions
to these shuffle relations.\\ \\
 In order to make this statement precise, we first need to extend cut-off and other regularised integrals on classical symbols  to cut-off and other regularised {\it iterated integrals} on tensor products of classical symbols; they are all  continuous linear
 forms on  spaces of symbols which naturally  extend to continuous linear
 forms on the (closed) tensor product. The Wodzicki residue, which is also
 continuous   on classical symbols of fixed order, extends in a similar way
 to a higher order residue density $\widetilde {{\rm res}}_{x,k}$ at point
 $x=(x_1,\ldots ,x_k)\in U=U_1\times\cdots\times U_k$ on the tensor product $\hat \otimes_{i=1}^kCS(U_i)$  and the well-known relation expressing the ordinary residue density ${\rm res}_x:= {\rm res}_{x, 0}$ as a complex residue:
$${\rm Res}_{z=0} \cutoffint_{T_x^*U}\sigma(z)(x,\xi)\, d\xi= -\frac{1}{\alpha^\prime(0)}{\rm res}_x(\sigma(0))\quad\forall \sigma\in CS(U)$$
extends to $\hat \otimes_{i=1}^kCS(U_i)$. Here $\sigma(z)$ is a holomorphic
family of classical symbols with order $\alpha(z)$ such that
$\alpha^\prime(0)\neq 0$.\\
 Indeed, the map $z\mapsto \cutoffint_{T_x^* U}\sigma(z)(x,\xi)\, d\xi $ with $\sigma\in \hat\otimes_{i=1}^k CS^{\alpha_i}(U_i)$ is  meromorphic  with poles of order no larger than $k$ and we have (see Theorem \ref{thm:altiteratedreg})
\begin{equation}\label{eq:resk}
{\rm Res}^{k}_{z=0} \cutoffint_{T_x^* U}\sigma(z)(x,\xi)\, d\xi= \frac{(-1)^k}{\prod_{i=1}^k\alpha_i^\prime(0)}\widetilde{{\rm res}}_{x,k}(\sigma(0))\quad\forall \sigma\in \hat \otimes_{i=1}^kCS(U_i),
\end{equation}
which is independent of the choice of regularisation ${\cal R}:\sigma \mapsto
\sigma(z)$ which sends the symbol
$\sigma$ to a holomorphic family of symbols 
$\sigma(z)$ such that $\sigma(0)=\sigma$. 
\\ \\
 Another approach to regularised {\it iterated} integrals is  to consider 
the operator  $
\sigma \mapsto  P(\sigma)$ 
$$ P(\sigma)(\eta)= \int_{\vert \xi \vert \leq \vert \eta\vert} \,\sigma(\xi)\, d\xi.$$
It maps $\sigma\in CS(U)$ to a symbol $P(\sigma)$ which
is not anymore classical, since it  raises the power of the logarithm entering
the asymptotic expansion of the symbol by one. The fact that  the algebra of
classical symbols is not stable under the action of $P$ justifies the
introduction of log-polyhomogeneous symbols in this context  (see e.g. \cite{L} for an extensive study of
log-polyhomogeneous symbols and operators).  Indeed, the operator $P$ satisfies a Rota-Baxter
relation (of weight zero):
$$P(\sigma)\, P(\tau)= P(\sigma\, P(\tau))+ P(\tau \, P(\sigma))$$ 
and defines a Rota-Baxter operator on the algebra of logpolyhomogeneous
symbols (see Proposition \ref{prop:RotaBaxter}). In one dimension the Rota-Baxter relation is an
integration by parts formula in disguise but for higher dimensions, this Rota
Baxter formula does not merely reduce to an integration by parts formula.
However,  similarities are to be expected between the obstructions to shuffle
relations  for regularised integrals studied here and the obstructions to
Stokes' formula for regularised integrals of symbol valued forms studied in
\cite{MMP}. In both cases the obstructions disappear under a non integrality
assumption on the orders of the
symbols involved. It is interesting to note that   regularised
integrals behave nicely specifically on symbols of non integer order, namely when they obey Stokes' property \cite{MMP} and  have good
transformation properties \cite{L}, \cite{MMP}.  \\
Unlike in the previous approach, we now take a fixed open subset $U\in \R^n$
so that $U_i=U, \, i=1, \cdots, k$. From a tensor product 
$\sigma= \sigma_1\otimes \cdots \otimes\sigma_k$ of classical symbols $\sigma_i\in CS(U)$ and operators  
\begin{eqnarray*}
\sigma&\mapsto & P_k(\sigma)\\
(\sigma)(x; \xi_1, \cdots, \xi_{k})&:=& P\left(\sigma(x;\xi_1,\cdots,\xi_{k},\cdot)\right)(\xi_{k}), 
\end{eqnarray*}
for fixed 
$x\in U$, one builds a map $(x, \xi)\mapsto \left(P_{1}\circ \cdots
  \circ P_{k-1}(\sigma)\right) ( x,\xi) $ which is
logpolyhomogeneous. The regularised  cut-off iterated integral of $\sigma$ can
then  be seen as an ordinary   regularised cut-off integral (extended by M. Lesch
\cite{L}  to
logpolyhomogeneous symbols)  on the logpolyhomogeneous symbol   $ P_{1}\circ \cdots
  \circ P_{k-1}(\sigma)$ in our case \footnote{ Similar nested integrals arise
    in D.Kreimer's work \cite{K1} in relation to a change of scale in the
    renormalisation procedure. His rooted trees describing nested integrations
  can  be adapted to our context, decorating trees with symbols
  $\sigma_i$. We thank D. Kreimer for pointing this reference out to us, which
  we read after this article was completed.}: 
$$\cutoffint_{ T_x^*U} \sigma= \sum_{\tau\in \Sigma_k}\cutoffint_{T_{x}^*
  U}  d\xi_1 \, P_{1}\circ \cdots \circ P_{k-1}\left(
  \sigma_\tau\right).$$
 \\
When $\sigma= \otimes \,\sigma_i$ and  the (left) partial sums
$\alpha_{1}+\alpha_{2}+\cdots +\alpha_{j}$, $j=1, \cdots, k$    of the orders $\alpha_i$ of the  symbols $\sigma_i \in CS(U)$ are {\it non integer}, the following shuffle relations hold (see Theorem \ref{thm:regiteratedRotaBaxter})
\begin{equation}\label{eq:shuffleintro}
\prod_{i=1}^k \cutoffint_{ T_x^*U} \sigma_i= \sum_{\tau\in \Sigma_k}\cutoffint_{T_{x}^* U}  d\xi_1 \, P_{1}\circ \cdots \circ P_{k-1}\left( \sigma_\tau\right)
\end{equation}
where we have set $\sigma_\tau:= \otimes_{i=1}^k \sigma_{\tau(i)}.$
\\
A holomorphic regularisation procedure ${\cal R}:\sigma \mapsto \sigma(z)$ on $ 
 CS(U)$ (with some continuity assumption)  induces a regularisation procedure
 $\sigma_1\otimes \cdots \otimes \sigma_k  \mapsto \sigma_1(z)\otimes \cdots
 \otimes \sigma_k(z)$   on $\hat\otimes^k CS(U)$. Using results by  Lesch  \cite{L} on cut-off integrals of holomorphic families of logpolyhomogeneous symbols  we build
 meromorphic maps
$z\mapsto \cutoffint_{ T_x^* U} \sigma(z)$ with poles of order at most $k$ for
any $\sigma \in \hat\otimes^k CS(U)$. \\ \\
When $\sigma(z)$ has order $q\, z+\alpha(0)$, then
equation  (\ref{eq:shuffleintro})
 implies the following   equality of
     meromorphic functions 
\begin{equation}\label{eq:shuffleintro2}
\prod_{i=1}^k \cutoffint_{ T_x^*U} \sigma_i(z)= \sum_{\tau\in \Sigma_k}\cutoffint_{T_{x}^* U}  d\xi \, P_{1}\circ \cdots \circ P_{k-1}\left( \otimes_{i=1}^k\sigma_{\tau(i)}(z)\right).
\end{equation}
But in general,  the constant term in the meromorphic expansion on the l.h.s
does not coincide with the product of the regularised integrals $\cutoffint_{T^*_xU}^{{\cal
R}}\sigma_i:=\hbox{fp}_{z=0}\cutoffint\sigma_i(z)$, namely in general
$$ {\rm fp}_{z=0}\prod_{i=1}^k \cutoffint_{ T_x^*U} \sigma_i(z)\neq
\prod_{i=1}^k \cutoffint_{ T_x^*U}^{\cal R} \sigma_i.$$  However, shuffle
relations extend to these regularised integrals  provided the symbols involved
have {\it vanishing Wodzicki residue} (see Corollary \ref{cor:renshuffle}):
$$\prod_{i=1}^k \cutoffint_{ T_x^*U}^{\cal R} \sigma_i= \sum_{\tau\in \Sigma_k}\cutoffint_{T_{x}^* U}^{\cal R}  d\xi_1 \, P_{1}\circ \cdots \circ P_{k-1}\left( \sigma_\tau\right).$$
For general symbols, a renormalisation procedure borrowed from  physicists
keeps track of counterterms one needs to introduce in order to pick the
``right'' finite part  thereby circumventing  the problem that ``taking finite parts'' does not commute with ''taking products'' of meromorphic functions. \\  \\  
The above constructions are adapted in section 5 to invariant
classical pseudodifferential operators acting on sections over the unit circle
$S^1$. Using the identification
$S^1\simeq \R/2\pi\Z$, one can relate the shuffle relations for integrals of the symbol of the modulus of
the Dirac operator on the circle  with ``second shuffle relations'' for
multiple zeta-functions. The adaptation is not straightforward as the symbol is
not a smooth function anymore; since it involves Dirac measures the integrals
turn out to be discrete sums.  The Euler-MacLaurin formula is the main tool
which enables us to go from integrals of symbols  to discrete
sums of symbols.\\ \\
These  shuffle relations for regularised integrals of symbols and their link with shuffle relations for zeta functions are a hint towards deeper algebraic  structures underlying  cut-off multiple integrals on one hand  and   renormalisation procedures in quantum field theory on the other hand. 
\\ In particular, this leads to the following open questions. Combining tensor
products $\sigma=\otimes_{i=1}^I\sigma_i$ considered previously  with injective linear maps
$B=B_0\otimes I_n : \R^L\otimes \R^n\to \R^I\otimes \R^n$ with $L\leq I$, one can build a class of functions\footnote{In the context of Feynman diagrams,  $L$ stands
  for the number of loops and $I$ for the number of internal edges. }  $$f(\xi_1
\cdots, \xi_L)=\sigma \circ B(\xi_1, \cdots, \xi_L)$$ in the momenta
$\xi_1,\cdots, \xi_k$ which, for certain choices of $\sigma_i$'s  are  of Feynman type in the language of Etingof
\cite{E}. A regularisation procedure $ {\cal R}$  on classical symbols gives
rise to holomorphic families $z\mapsto \sigma_i(z)$ from which we can build a
map $ (z_1, \cdots, z_I) \mapsto  \sigma_{z_1, \cdots,
z_I}=\otimes_{i=1}^I\sigma_i(z_i)$. It  is reasonable to  expect the map 
$$ (z_1, \cdots, z_I) \mapsto \cutoffint \sigma _{z_1,\cdots,z_I}\circ
B(\xi_1, \cdots, \xi_L)\,d\xi_1\cdots d\xi_L $$
to give rise to a Laurent expansion in the $z_i$'s, on the grounds of work by Speer
\cite{S}\footnote{We thank Dirk Kreimer for drawing our attention to this
  reference. Speer's
    results are transposed here to the euclidean set up.}
who proves this fact when $\sigma_i(\xi)= (\vert \xi\vert^2+m_i^2)^{-1}\quad
\forall i\in\{1, \cdots, I\}$  and $\sigma(z)=\sigma^{1+z}$.  Alternatively, following
a dimensional  regularisation type procedure, one can  build maps 
$$ (z_1,\cdots, z_L)\mapsto \int \sigma \circ B(\xi_1,\cdots, \xi_L)\, \vert
\xi_1\vert^{-z_1}\cdots \vert\xi_L\vert^{-z_L}\, d\xi_1\cdots d\xi_L, $$
which again can be expected to give rise to Laurent expansions  and hence to 
 a  meromorphic function  at $0$ when  $z_1=\cdots =z_L=z$.  Etingof's results
on dimensional regularisation \cite{E} imply 
this meromorphicity property   when $\sigma_i(\xi)= (\vert \xi\vert^2+m_i^2)^{-1}\quad
\forall i\in \{1, \cdots, I\}$  on the grounds of   a theorem by Bernstein but
 further investigations are   needed  to prove the first part of the statement on the existence of
a Laurent expansion in several
variables. 
\\
In Theorem \ref{thm:regprod}, we show that provided the class of functions under consideration has
this expected  Laurent expansion behaviour,
the renormalisation procedure  boils down to picking up   the constant term in
the  Laurent expansion in $(z_1, \cdots, z_k)$.  This fact was already proved
by Speer \cite{S} in the particular case  we briefly
described above in relation to his work.\\ \\
It appears from the  investigations carried out here,  that iterated integrals of symbols
seem to  provide a stepping stone between Feynman type integrals in physics
and the renormalisation procedures used to handle their divergences on one hand and
multiple zeta functions and the regularised shuffle relations they obey, a line of
thought we want to pursue further in a forthcoming paper. 
\vfill \eject \noindent 
\section{Preliminaries}
 For $\alpha\in \R$, $k\in \N$, the set    $CS^{\alpha, k}(U)$  of scalar valued logpolyhomogeneous  symbols of order $\alpha$ on  an open subset $U$ of $\R^n$   can be equipped  with a Fr\'echet  structure. Such a symbol reads:
\begin{equation}\label{eq:localsymb}
\sigma= \sum_{m=0}^{N-1}\psi\, \sigma_{\alpha-m}+ \sigma_{(N)},
\end{equation}
where $\psi$ is a smooth function which vanishes at $0$ and equals to one outside a compact, where  $ \sigma_{\alpha-m}(x, \xi)=\sum_{p=0}^k \sigma_{\alpha-m, p}(x, \xi)\, \log^p \vert \xi\vert \in \Ci(S^*U)$ with $ \sigma_{\alpha-m, p}(x, \xi)$  positively homogeneous in $\xi$ of order $\alpha-m$ and where $\sigma_{(N)}\in \Ci(S^*U)$ is a symbol of order $\alpha-N$. 
The following semi-norms labelled by multiindices $\gamma,\beta$ and integers $m\geq 0, p\in \{1, \cdots, k\}$, $N$ give rise to a Fr\'echet topology on
 $CS^{\alpha,k}(U)$:
\begin{eqnarray*}
&{} & {\rm sup}_{x\in K, \xi \in \R^n} (1+\vert \xi\vert)^{-\alpha+\vert \beta\vert} \, \vert \partial_x^\gamma \partial_\xi^\beta \sigma(x, \xi)\vert;\\
&{}&  {\rm sup}_{x\in K, \xi\in \R^n}  \vert \xi\vert^{-\alpha+N+\vert
  \beta\vert}\,\vert \partial_x^{\gamma} \partial_\xi^{\beta}
\left(\sigma-\sum_{m=0}^{N-1} \psi(\xi)\, \sigma_{\alpha-m}\right)(x, \xi) \vert;\\
&{}& {\rm sup}_{x\in K, \vert\xi\vert=1}  
\vert \partial_x^{\gamma} \partial_\xi^{\beta} \sigma_{\alpha-m, p}(x, \xi) \vert,
\end{eqnarray*}
where $K$  ranges over  compact sets in $U$. 
\begin{rk}
Note that the first set of norms corresponds to the ordinary symbol topology, the second set of norms controls the rest term $\sigma_{(N)}$ whereas the last set of norms is the ordinary supremum norm on the homogeneous components of the symbol. 
\end{rk}
Let us introduce some notations.
The set  $CS^{-\infty}(U):= \bigcap_{m\in \R} CS^m(U)$ corresponds to  the algebra of smoothing symbols. 
The set $$CS^{\Z,*}(U):= \bigcup_{m\in \Z}\bigcup_{k\in \N} CS^{m,k}(U)$$
of integer order log-polyhomogeneous symbols,
which is equipped with an  inductive limit topology of Fr\'echet spaces is
strictly contained in the algebra generated by log-polyhomogeneous symbols of
any order 
 $$CS^{*,*}(U):= \langle \bigcup_{m\in \R}\bigcup_{k\in \N} CS^{m,k}(U)\rangle.$$
 
Following \cite{KV} (see also \cite{L}), we   extend the continuity on symbols of fixed order to families of symbols with varying order as follows:
\begin{defn}\label{defn:1}
Let $k$ be a non negative integer.
A map  $b\mapsto \sigma(b)\in CS^{*, k}(U)$ of symbols parametrized by a topological space $B$ is continuous     if the following
 assumptions hold:
 \begin{enumerate}
 \item the order $\alpha(b)$ of $\sigma(b)$
is continuous  in $b$,
\item  for any non negative integer $j$, the homogeneous components
$\sigma_{ \alpha(b) -j, l} (b)(x, \xi)$, $0\leq l\leq k$ of the symbol $\sigma( b)(x, \xi) $
yield  continuous  maps $b\mapsto \sigma_{ \alpha(b) -j} (b):=\sum_{l=0}^k \sigma_{ \alpha(b) -j, l}\log^l\vert \xi\vert $ into  $\Ci(T^*U )$,
\item   for any sufficiently large integer $N$, the truncated kernel $$K^{(N)}(b)(x, y):= \int_{T_x^*U}d\xi e^{i \xi\cdot (x-y)}  \sigma_{(N)}(b)(x, \xi)$$ where  $$\sigma_{(N)}(b)(x, \xi):= \sigma(b)(x,\xi)- \sum_{j=0}^N \psi(\xi) \, \sigma_{\alpha(b)-j}(b)(x, \xi)$$
 yields  a continuous  map $b\mapsto \sigma_{(N)}(b)$  into  some  $C^{K(N)} (U\times U)$  where $\lim_{N\to \infty} K(N)=+\infty$.
\end{enumerate}
\end{defn}
\section{Regularised  integrals of log-polyhomogeneous  symbols}
We recall for completeness,  well-known results on regularisation techniques of integrals of ordinary log-polyhomogeneous  symbols which lead to trace functionals on the corresponding pseudodifferential operators. 
\subsection{Cut-off integrals of log-polyhomogeneous  symbols}
We start by recalling the construction of cut-off integrals of log-polyhomogeneous symbols  \cite{L} which generalizes results previously  established by Guillemin and Wodzicki in the case of classical symbols.
\begin{lem}\label{lem:cutoff}
Let $U$ be an open subset of $\R^n$ and  for any non negative integer $k$, let $\sigma\in CS^{*,k}(U)$ be a log-polyhomogeneous symbol, then for any $x\in U$
\begin{itemize}
\item $\int_{B_x^*(0, R)} \sigma(x,\xi) d\xi$ has an asymptotic
expansion in $R\to\infty$ of  the form:
\begin{eqnarray}\label{eq:cutoffasympt}
\int_{B_x^*(0, R)} \sigma(x,\xi) d\xi&\sim_{ R\to \infty}&C_x(\sigma)+ \sum_{j=0,\alpha-j+n\neq 0}^\infty \sum_{l=0}^k P_l(\sigma_{\alpha-j, l})(\log R)  \, R^{\alpha-j+n}\nonumber\\
&+&
\sum_{l=0}^k \frac{{\rm res}_{x,l}(\sigma)}{l+1}\log^{l+1}  R
\end{eqnarray} where $P_l(\sigma_{\alpha-j, l})(X)$ is a polynomial of degree $l$ with coefficients depending on $\sigma_{\alpha-j, l}$ and where  $C_x(\sigma)$ is the constant term corresponding to the finite part:
\begin{eqnarray*}
C_x(\sigma) &
:=& \int_{T_x^*U}\sigma_{(N)}(x, \xi)\, d\xi+  \int_{B_x^*(0, 1)}
 \psi(\xi) \sigma(x, \xi)\, d\xi\\
 &+&\sum_{j=0, \alpha-j+n\neq 0}^{N}\sum_{l=0}^k \frac{(-1)^{l+1}l!}{(\alpha-j+n)^{l+1}}
 \int_{S_x^*U} \sigma_{\alpha-j,l} (x,\xi) d_S\xi\\
\end{eqnarray*}

which is independent of   $N\geq \alpha +n-1$.

\item For any fixed $\mu>0$,
$${\rm fp}_{R\to\infty}\int_{B_x^*(0,\mu \, R)}\sigma(x,\xi) d\xi={\rm fp}_{R\to\infty} \int_{B_x^*(0, R)} \sigma(x,\xi)\, d\xi+\sum_{l=0}^k \frac{\log^{l+1}\mu}{l+1} \cdot{\rm res}_{l,x}(\sigma).$$
\end{itemize}
\end{lem}

\begin{rk}
If $\sigma$ is a classical operator, setting $k=0$
in the above formula yields
\begin{eqnarray*}
{\rm fp}_{R\to\infty}\int_{B_x^*(0, R)} \sigma(x,\xi) d\xi
 &:=& \int_{T_x^*U}\sigma_{(N)}(x, \xi) \, d\xi+\sum_{j=0}^{N}  \int_{B_x^*(0, 1)}
 \psi(\xi)\, \sigma_{\alpha-j}(x,\xi) \, d\xi\\
&- &\sum_{j=0, \alpha-j+n\neq 0}^{N}  \frac{ 1}{\alpha-j+n}
 \int_{S_x^*U} \sigma_{\alpha-j } (x,\omega) d\omega.
\end{eqnarray*}
\end{rk}
{\bf Proof.}
Given a log-polyhomogeneous symbol $\sigma\in CS^{\alpha, *}(U)$,  for   any $N\in \N$ we write:
\begin{equation}
\sigma (x,
\xi)=
\sum_{j=0}^N \psi (\xi) \sigma_{\alpha-j}(x,\xi)+
\sigma_{(N)}(x, \xi)\quad \forall  (x, \xi)\in
T^*U,
\end{equation}
where $\sigma_{(N)}\in S^{\alpha-N-1}(U)$.
\begin{itemize}
\item For some fixed $N\in \N$ chosen large enough such that $\alpha-N-1<-n$, we  write $\sigma(x,\xi)=
\sum_{j=0}^{N} \psi \sigma_{\alpha-j} (x,\xi)+\sigma_{(N)}(x,\xi)$
and split the integral accordingly:
$$\int_{B_x^*(0, R)}  \sigma(x,\xi) d\xi= \sum_{j=0}^{N}\int_{B_x^*(0, R)}\psi(\xi) \sigma_{\alpha-j}(x,\xi) d\xi+ \int_{B_x^*(0, R)} \sigma_{(N)}(x,\xi) d\xi.$$
Since $\alpha-N-1<-n$,  $\sigma_{(N)}$ lies in $ L^1(T_x^*U)$ and the
integral  $\int_{B_x^*(0, R)} \sigma_{(N)}(x,\xi) d\xi$ converges when
$R\to \infty$ to $\int_{T_x^*U} \sigma_{(N)}(x,\xi) d\xi$. On the other
hand, for any $j\leq N$
\begin{eqnarray}\label{eq:intlambda}
\int_{B_x^*(0, R)} \psi(\xi) \sigma_{\alpha-j}(x, \xi)= \int_{B_x^*(0, 1)} \psi
(\xi)\sigma_{\alpha-j}(x, \xi)+ \int_{D_x^*(1,R)} \sigma_{\alpha-j}(x, \xi)
\end{eqnarray}
 since $\psi$ is constant equal to $1$ outside the unit ball. Here $D_x^*(1, R)=
 B_x^*(0,R)-B_x^*(0, 1)$. The first integral on the r.h.s.  converges and
  since  $$\sigma_{\alpha-j}(x,\xi)=\sum_{l=0}^k \sigma_{\alpha-j, l}(x,\xi)\log^l \vert \xi\vert,$$  the second integral reads:
$$\int_{D_x^*(1,R)} \sigma_{\alpha-j}(x,\xi) d\xi= \sum_{l=0}^k\int_1^R  r^{\alpha-j+n-1}\log^l r \, dr\cdot \int_{S_x^*U} \sigma_{\alpha-j,l}(x, \omega)d\omega.$$
Hence the following asymptotic behaviours:
\begin{eqnarray*}
 &{}& \int_{D_x^*(1,R)} d\xi\, \sigma_{\alpha-j}(x,\xi)\\
&\sim_{R\to \infty}&
 \sum_{l=0}^k\frac{ \log^{l+1}R}{l+1}\cdot \int_{S_x^*U}
\sigma_{\alpha-j,l}( x,\omega)d\omega=\sum_{l=0}^k
\frac{\log^{l+1}R}{l+1} {\rm res}_{l,x}(\sigma)\quad \, {\rm if}
\, \,  \alpha-j=-n
\end{eqnarray*}
whereas:\eject \noindent
\begin{eqnarray*}
 \int_{D_x^*(1,R)} \sigma_{\alpha-j}&\sim_{R\to \infty}&   \sum_{l=0}^k
 \left(\sum_{i=0}^l\frac{(-1)^{i+1}\frac{l!}{(l-i)!}\log^i R}{(\alpha-j+n)^i}\cdot R^{\alpha-j+n}\int_{S_x^*U} \sigma_{\alpha-j,l}(x, \omega)d\omega\right.\\
&+&
(-1)^{l}l!\frac{R^{\alpha-j+n}}{(\alpha-j+n)^{l+1}}\cdot \int_{S_x^*U} \sigma_{\alpha-j,l} (x,\omega) d\omega \\
&+& \left. \frac{(-1)^{l+1}l!}{(\alpha-j+n)^{l+1}}\cdot \int_{S_x^*U} \sigma_{\alpha-j,l} (x,\omega) d\omega\right)\, {\rm if}\quad \, \, \alpha-j\neq -n.\\
\end{eqnarray*}
Putting together these asymptotic expansions  yields the statement
of the proposition with
$$C_x (\sigma)= \int_{T_x^*U}\sigma_{(N)}+\sum_{j=0}^{N}  \int_{B_x^*(0, 1)}
 \psi \sigma_{a_j}+ \sum_{j=0, a_j+n\neq 0}^{N}\sum_{l=0}^L \frac{(-1)^{l+1}l!}{(a_j+n)^{l+1}}\int_{S_x^*U} \sigma_{a_j,l}.$$
\item The $\mu$-dependence follows from
\begin{eqnarray*}
\log^{l+1}(\mu \, R)&=&\log^{l+1}R\, \left(1+\frac{\log\mu}{\log \, R}\right)^{l+1}\\
&\sim_{R\to\infty}&\log^{l+1}R\,\sum_{k=0}^{l+1} C_{l+1}^k \left(\frac{\log\mu}{\log R}\right)^k.\\
\end{eqnarray*}
The logarithmic terms $\sum_{l=0}^k \frac{{\rm
res}_{l,x}(\sigma)}{l+1}\log^{l+1}(\mu\, R)$  therefore contribute
to the finite part by $\sum_{l=0}^k \frac{\log^{l+1}\mu}{l+1}\cdot
{\rm res}_{l,x}(\sigma)$ as claimed in the lemma.\endsquare

\end{itemize}

Discarding the divergences, we can therefore  extract a finite part from the
asymptotic expansion of $\int_{B(0, R)}\sigma(x,\xi) d\xi$ and
set for $\sigma\in CS^{*,k}(\R^n)$:
\begin{defn}
Given an non negative integer $k$, an open subset $U\subset \R^n$  and a point $x\in U$, for  any $\sigma\in CS^{\alpha, k}(U)$, the cut-off integral
\begin{eqnarray}\label{eq:cutoffint}
 &{}&\cutoffint_{T_x^*U} \sigma(x,\xi) d\xi :={\rm fp}_{R\to\infty } \int_{B_x^*(0,R)} \sigma(x,\xi) d\xi\nonumber\\
&= & \int_{T_x^*U}\sigma_{(N)}(x, \xi) \, d\xi +\sum_{j=0}^{N}  \int_{B_x^*(0, 1)}
 \psi(\xi) \sigma_{\alpha-j}(x, \xi) \, d\xi\nonumber\\
&+& \sum_{j=0, \alpha-j+n\neq 0}^{N}\sum_{l=0}^k \frac{(-1)^{l+1}l!}{(\alpha-j+n)^{l+1}}\int_{S_x^*U} \sigma_{\alpha-j,l}(x, \xi)\, d_S\xi
\end{eqnarray}
is independent of $N> \alpha+n-1$.\\
It is independent of the parametrisation $R$ provided the higher Wodzicki residue
$${\rm res}_{x,l}:= \int_{S_x^*U} \sigma_{-n,l } (x,\xi) d_S\xi$$ vanishes for all integer $0\leq l\leq k$.
\end{defn}
This explicit description of the finite part leads to the following continuity result. 
\begin{prop}\label{prop:contcutoff}
For any fixed $\alpha\in \R$ and any non negative integer $k$, and given an open subset $U\in \R^n$, a point $x\in U$,  the map 
\begin{eqnarray*} 
CS^{\alpha, k}(U)&\to & \Ci(U,\C)\\
\sigma&\mapsto& \left(x\mapsto \cutoffint_{T_x^*U}\sigma(x, \xi) \, d\xi\right)\\
\end{eqnarray*}
is continuous in the  Fr\'echet topology of $CS^{\alpha, k}(U)$ and the natural  topology of $\Ci(U, \C)$. 
\end{prop}
\begin{rk}The assumption that $\alpha$ be fixed is essential here. 
\end{rk}
{\bf Proof:} From formula (\ref{eq:cutoffint}) and the fact that symbols are
smooth functions on $U\times \R^n$, it follows that the cut-off integral is
$\Ci(U, \C)$-valued. \\
The maps $\sigma\mapsto \left(x\mapsto \int_{B_x^*(0, 1)}
 \psi(\xi) \sigma_{\alpha-j}(x, \xi) \, d\xi\right)$ and $\sigma\mapsto\left(x\mapsto  \int_{S_x^*U} \sigma_{\alpha-j,l}(x, \xi)\, d_S\xi\right)$ are clearly continous as integrals over compact sets of continuous maps. On the other hand the map $\sigma\mapsto \left(x\mapsto \int_{T_x^*U}\sigma_{(N)}(x, \xi) \, d\xi \right)$ is continuous since $\sigma\mapsto \sigma_{(N)}$ is continuous and $\sigma_{(N)}(x, \xi)\leq C (1+\vert  \xi \vert)^{-N}$ can be uniformly bounded by an $L^1$ function.\endsquare \\ \\
As well as the higher order residue density function ${\rm res}_{x, k}$, one can define  on  $CS^{*, k}(U)$ an extension of 
the  ordinary residue density function ${\rm res}_x$ as follows:
$${\rm res}_x(\sigma):= \int_{S_x^*U}\left(\sigma(x, \xi)\right)_{-n}\, d_S\xi$$ 
where $d_S\xi$ is the volume measure on the unit cotangent sphere $S_x^*U$
induced by the canonical volume measure on $T_x^*U$. Even though it certainly
does not induce a graded trace on the algebra of log-polyhomogeneous operators
on a closed manifold as the higher order residue does \cite{L}, it is a useful
tool for what follows since  we have the following  continuity result:
\begin{lem}\label{lem:contres}
Given any non negative integer $k$, and given any $\alpha\in \R$,  the map:
\begin{eqnarray*}
CS^{\alpha, k}(U)&\to & \Ci(U, \C)\\
\sigma &\mapsto &\left( x\mapsto {\rm res}_x(\sigma)\right)\\
\end{eqnarray*}
is continuous for the Fr\'echet topology on $CS^{\alpha, k}(U)$. 
\end{lem}
\subsection{Integrals of holomorphic families of log-polyhomogenous symbols}
Following \cite{KV} (see also \cite{L}), we   define a holomorphic family of logpolyhomogeneous symbols in $CS^{*,k}(U)$ in the same way as in definition \ref{defn:1} replacing continuous by holomorphic. \\
We quote from \cite{PS}  the following theorem which extends results of
\cite{L} relating the Wodzicki residue of holomorphic families of
log-polyhomogeneous symbols with higher  Wodzicki residues. For simplicity, we
restrict ourselves to holomorphic  families with order $\alpha(z)$ given by an
affine function of $z$, a case which covers natural applications. 
 \begin{thm}\label{thm:KV} Let $U$ be an open subset of $\R^n$  and let $k$ be a non negative integer. For any holomorphic 
family  $z\mapsto\sigma(z)\in CS^{\alpha(z), k}(U)$ of
symbols parametrised by  a domain  $W\subset \C$
such that $z\mapsto \alpha(z)=\alpha^\prime(0)\, z+ \alpha(0)$ is  an affine function with  $\alpha^\prime(z)=\alpha^\prime(0)\neq 0 $, then for any $x\in U$,  there is a Laurent expansion in a neighborhood of any $z_0\in P$
\begin{eqnarray*}
\cutoffint_{T_x^*U}
\sigma(z)(x,
\xi) d\xi &=& {\rm fp}_{z=z_0}\cutoffint_{T_x^*U}
\sigma(z)(x,
\xi) d\xi\\
&+& \sum_{j=1}^{k+1} \frac{r_j(\sigma)(z_0) (x)}{(z-z_0)^{j}} \\
&+& \sum_{j=1}^{K} s_j(\sigma)(z_0) (x)\, (z-z_0)^{j}\\
 &+& o\left((z-z_0)^K\right),\\
\end{eqnarray*}
 where for $1\leq j\leq k+1$,
$R_j(\sigma)(z_0) (x)$ is locally  explicitly determined by a local expression (see \cite{L} for the case $\alpha^\prime(0)=1$)
\begin{eqnarray} \label{eq:KVlogsymbresj}
&{}& r_j(\sigma)(z_0) (x)\nonumber\\
&:=&\sum_{l=j-1}^k \frac{(-1)^{l+1}}{(\alpha^\prime(z_0))^{l+1}}\frac{l!}{(l+1-j)!}\,  \,{\rm res}_x \left( \left(\sigma_{( l)}\right)^{(l+1-j)}\right) (z_0).
\end{eqnarray}
Here $\sigma_{(l)}(z)$ is the local symbol given by the coefficient of $\log^l \vert \xi\vert$ of $\sigma$ i.e. $$\sigma(z)= \sum_{l=0}^k \sigma_{(l)}(z)\log^l \vert \xi\vert.$$
On the other hand,  the finite part ${\rm fp}_{z=z_0}\cutoffint_{T_x^*U}
\sigma(z)(x,
\xi) d\xi $ consists of a global piece $\cutoffint_{\R^n} \sigma(z_0)(x, \xi)\, d\xi$ and a local piece:
\begin{eqnarray} \label{eq:KVlogsymb0}
{\rm fp}_{z=z_0}\cutoffint_{T_x^*U}
\sigma(z)(x,
\xi) d\xi& = &\cutoffint_{T_x* U} \sigma(z_0)(x, \xi)\, d\xi \nonumber\\
&+& \sum_{l=0}^k \frac{(-1)^{l+1}}{(\alpha^\prime(z_0))^{l+1}}\frac{1}{l+1}\,  \,{\rm res}_x \left( \left(\sigma_{( l)}\right)^{(l+1)}\right) (z_0).
\end{eqnarray}
Finally, for $1\leq j\leq K$, $S_j(\sigma)(z_0)(x)$ reads
\begin{eqnarray}\label{eq:KVlogsymbj}
 s_j(\sigma)(z_0)
 &:= &\cutoffint_{T_x^*U}\sigma^{(j)}(z_0) \, d\xi\nonumber\\
&+&\sum_{l=0}^k\frac{(-1)^{l+1}l!j!}{(\alpha^\prime(z_0))^{l+1}\,(j+l+1)!}
\,{\rm res}_x\left(  \left(\sigma_{(l)}\right)^{(j+l+1)}(z_0) \right). 
\end{eqnarray}
\end{thm}
As a consequence, the finite part ${\rm fp}_{z=z_0}\cutoffint_{T_x^*U}
\sigma(z)(x,
\xi) d\xi$ is entirely determined by the derivative $\alpha^\prime(z_0)$ of the order and by the derivatives of the symbol  $\sigma^{(l)}(z_0), \quad l\leq k+1$  via the cut-off  integral and the Wodzicki residue density. 
\subsection{Regularised integrals of log-polyhomogeneous symbols}
    Let us briefly recall the  notion of holomorphic regularisation taken from
    \cite{KV} (see also \cite{PS}).                                   
\begin{defn}
A  holomorphic regularisation procedure on $CS^{*, k}(U)$ for any fixed non negative integer $k$ 
is a  map 
\begin{eqnarray*}
{\cal R}: CS^{*, k}(U)&\to & {\rm Hol}\,\left( CS^{*, k}(U)\right)\\
\sigma  &\mapsto & \sigma(z)
\end{eqnarray*}
where ${\rm Hol}\left( CS^{*, k}(U)\right)$ is the algebra of holomorphic maps with values in $CS^{*, k}(U)$, 
such that 
\begin{enumerate}
\item  $\sigma(0)=\sigma$, 
\item  $\sigma(z)$ has holomorphic  order  $\alpha(z)$ (in particular, $\alpha(0)$ is equal to the order of $\sigma$)  such that $\alpha^\prime(0)\neq 0$.\\
 \end{enumerate}
We call a regularisation procedure  ${\cal R}$ continuous whenever   the map 
\begin{eqnarray*}
{\cal R}: CS^{*, k}(U)&\to &{\rm Hol}\left( CS^{*, k}(U)\right)\\
\sigma &\mapsto&\left( z\mapsto \sigma(z)\right)\\
\end{eqnarray*}
  is continuous. \\ \\
\end{defn}
\begin{rk} It is easy to check \cite{PS} that if  $z\to \sigma(z)\in CS^{\alpha(z), k}(U)$  then $\sigma^{(j)}(z_0)\in CS^{\alpha(z_0), k+j}(U)$. 
\end{rk}
Examples of holomorphic regularisations are the well known Riesz
regularisation $\sigma\mapsto \sigma(z)(x, \xi):= \sigma(x, \xi)\cdot \vert
\xi \vert^{-z}$ and generalisations  of the type  $\sigma\mapsto \sigma(z)(x,
\xi):=H(z)\cdot \sigma(x, \xi)\cdot \vert \xi \vert^{-z}$ where $H$ is a
holomorphic function such that $H(0)=1$. The latter  include dimensional
regularisation (see \cite{P}). These regularisation procedures are clearly
continuous. \\ As a consequence of the results of the previous paragraph, given
a holomorphic regularisation procedure ${\cal R}:\sigma \mapsto \sigma(z)$ on
$CS^{*, k}(U)$  and a symbol $\sigma\in CS^{*, k}(U)$, for every point $x\in
U$,  the map $z\mapsto \cutoffint_{T_x^*U} \,  \sigma(z)(x,\xi) \, d\xi$  is meromorphic with  poles of order at most $k+1$ at points in $\alpha^{-1}([-n, +\infty[\, \cap\, \Z)$ where $\alpha$ is the order of $\sigma(z)$ so that we can define the finite part when $z\to 0$   as follows. 
\begin{defn}
Given  a holomorphic regularisation procedure ${\cal R}:\sigma \mapsto \sigma(z)$ on $CS^{*, k}(U)$, a symbol $\sigma\in CS^{*, k}(U)$ and any point $x\in U$, we define the regularised integral 
\begin{eqnarray*}
\int_{T_x^*U}^{{\cal R}} \sigma(x, \xi) \, d\xi &:=& {\rm fp}_{z=0} \cutoffint_{T_x^*U} \sigma(z)(x, \xi)\, d\xi\\
&:=& \lim_{z\to 0}\left(  \cutoffint_{T_x^*U}d\xi \,  \sigma(z)(x, \xi)- \sum_{j=1}^{k+1}
\frac{1}{z^j} {\rm Res}^j_{z=0}\cutoffint_{T_x^*U} d\xi\, \sigma(z)(x, \xi)\right).
\end{eqnarray*}
\end{defn}
We have the following continuity result. 
\begin{prop}\label{prop:contregint}
Given  a continuous holomorphic regularisation procedure ${\cal R}:\sigma \mapsto \sigma(z)$ on $CS^{*, k}(U)$ where $k$ is a non negative integer, for any fixed $\alpha\in \R$, there is a discrete set $P_\alpha\subset \C$  such that the map
\begin{eqnarray*} 
CS^{\alpha, k}(U)&\to &\Ci(U, {\rm Hol}( \C-P_\alpha))\\
\sigma&\mapsto& \cutoffint_{T_x^*U}\sigma(x, \xi)(z) \, d\xi\\
\end{eqnarray*}
is continuous  on $\Ci \left(U, {\rm Hol}( \C-P_\alpha)\right)$.
Moreover the map
\begin{eqnarray*} 
CS^{\alpha, k}(U)&\to &\Ci(U,  \C)\\
\sigma&\mapsto& \cutoffint_{T_x^*U}^{{\cal R}}\sigma(x, \xi) \, d\xi\\
\end{eqnarray*}
is continuous on  $CS^{\alpha, k}(U)$. 
\end{prop}
\begin{rk}The assumption that $\alpha$ be constant is essential here. 
\end{rk}
{\bf Proof:} From Theorem \ref{thm:KV} we know that the map $z\mapsto\cutoffint_{T_x^* U}\sigma(z)(x,\cdot)$ is meromorphic with simple poles in some discrete set $P_\alpha$.  From Proposition 1 we know that the map $\sigma\mapsto\cutoffint \sigma$ is continuous. Combining these two results gives the continuity of the map $ \sigma\mapsto \left(z\mapsto  \cutoffint_{T_x^*U}\sigma(x, \xi)(z) \, d\xi\right)$ where the r.h.s is understood as a holomorphic map on $\C-P_\alpha$. \\
We now prove the second part of the proposition. By theorem \ref{thm:KV}  applied to $z_0=0$, it is sufficient to check that the maps
$\sigma\mapsto  \cutoffint_{T_x^*U} \sigma(0)(x, \xi) \, d\xi$ and 
the maps $\sigma\mapsto {\rm res}_x\left(\sigma^{(j)}(0)\right)$ are $\Ci(U, \C)$ valued and continuous for any $1\leq j\leq k+1$ for the Fr\'echet topology on log-polyhomogenous symbols and the Fr\'echet topology on smooth functions.\\
 From the continuity assumption on the regularisation ${\cal R}$ combined with Proposition \ref{prop:contcutoff} and Lemma \ref{lem:contres} it follows that  for a log-polyhomogeneous symbol $\tau$, both $x\mapsto \cutoffint_{T_x^*U}\tau(x, \xi)\, d\xi$ and $x\mapsto {\rm res}_x(\tau)$ are smooth functions. Applying this to $\tau=  \sigma^{(j)}(0)$ (which is log-polyhomogeneous by the above remark) with $0\leq j\leq k+1$ yields the result. \endsquare
\section{ Regularised integrals on tensor products of classical symbols}
\subsection{Tensor products of  symbols}
Let $U_1, \cdots, U_L$ be open subsets of $\R^n$. 
Since the spaces $CS^{m_i}(U_i)$ and $CS^{m_i, k_i}(U_i)$ are Fr\'echet spaces, we can form their closed tensor products, where the closed tensor product of two Fr\'echet spaces $E$ and $F$ is the Fr\'echet space $E\hat \otimes F$ built as the closure of $E\otimes F$ for the finest topology for which $\otimes: E\times F\to E\otimes F$ is continuous. \\
\begin{defn} For any multiindices $(m_1, \cdots, m_L)\in R^L$,  $(k_1, \cdots, k_L)\in \N^L$ we set
$$  CS_{w}^{(m_1, \cdots, m_L)}\left(U_1\times \cdots \times U_L\right):= \hat\otimes_{i=1}^L CS^{m_i}\left(U_i\right)$$ and
$$  CS_{w}^{(m_1, \cdots, m_L)\, , (k_1, \cdots, k_L)}\left(U_1\times \cdots \times U_L\right):= \hat\otimes_{i=1}^L CS^{m_i, k_i}\left(U_i\right).$$
The multiindex $(m_1, \cdots, m_L)$ is called the multiple order of $\sigma$ and $m_1+\cdots+ m_L$ its total order. \\
\end{defn}
There are at least two ways of continuously  extending regularised integrals to  tensor products of symbols.
\subsection{A first  extension of regularised integrals to tensor products}
\begin{defn} Let $U=U_1\times \cdots \times U_L$ with $x=(x_1,\cdots, x_L)$, $x_i\in U_i, i=1, \cdots, L$ open subsets in $\R^n$. Let $(\alpha_1,\cdots, \alpha_L)\in \C^l$ and let $(k_1, \cdots, k_L)$ be a multiindex of non negative integers. \\ 
The continuous  maps 
\begin{eqnarray*} 
CS^{\alpha_i, k_i}(U_i)&\to & \Ci(U_i, \C)\\
\sigma_i&\mapsto&\left(x_i\mapsto  \cutoffint_{T_{x_i}^*U_i}\sigma_i(x_i, \xi_i)\, d\xi_i\right),\quad i=1, \cdots, L\\
\end{eqnarray*}
induce a uniquely defined map:
\begin{eqnarray*} 
CS_w^{(\alpha_1, \cdots, \alpha_L), (k_1, \cdots, k_L)}(U)&\to & \Ci(U, \C)\\
\sigma &\mapsto&\left(x\mapsto  \altcutoffint_{T_{x}^*U}\sigma(x, \xi)\,d\xi_1\cdots d\xi_L\right)\\
\end{eqnarray*}
which gives rise to a linear map on $\hat\otimes^k CS(U_i)$
called the multiple regularised cut-off integral of $\sigma (x,\cdot)$. 
\end{defn}
Clearly, if $\sigma(x,\cdot)=\otimes_{i=1}^k \sigma_i\sigma_i(x_i,\cdot)$ we have: 
$$ \altcutoffint_{T_{x}^*U}\sigma(x, \xi)\,d\xi_1\cdots d\xi_L= \prod_{i=1}^L
 \altcutoffint_{T_{x_i}^*U}\sigma(x_i, \xi_i)\,d\xi_i.$$ 
The following extends holomorphic regularisations to tensor products of symbol spaces. 
\begin{defn} Let $U=U_1\times \cdots \times U_L$ be a product of open subsets of $\R^n$. 
For a given multiindex $(k_1, \cdots, k_L)$ with $k_i$ non negative integers, a regularisation procedure ${\cal R}$ on $CS_w^{*,(k_1, \cdots, k_L)}(U)$ is a map:
\begin{eqnarray*}
{\cal R}: CS_w^{*, (k_1, \cdots, k_L)}(U)&\to & {\rm Hol}\left(CS_w^{*, (k_1, \cdots, k_L)}(U)\right)\\
\sigma&\mapsto & {\cal R}(\sigma):z\mapsto \sigma(z)\\
\end{eqnarray*}
such that\begin{enumerate}
\item  $\sigma(0)=\sigma$, 
\item  $\sigma(z)$ has holomorphic (multiple) order  $\alpha(z)=(\alpha_1(z),\cdots, \alpha_L(z))\in \R^L$ (in particular, $\alpha(0)$ is equal to the (multiple) order of $\sigma$)  such that ${\rm Re}\left(\alpha_i^\prime(0)\right)> 0$ for all $i\in \{1, \cdots, L\}$.\\
 \end{enumerate} 
Here ${\rm Hol}\left( CS_w^{*, (k_1, \cdots, k_L)}(U)\right)$ is the algebra of holomorphic maps with values in $CS^{*, k}(U)$.
\end{defn}
Clearly, regularisation procedures ${\cal R}_1, \cdots, {\cal R}_L$ on $CS^{*, k_1}(U_1), \cdots, CS^{*, k_L}(U_L)$ induce a regularisation procedure 
${\cal R}= \hat \otimes_{i=1}^L {\cal R}_i$ on $CS_w^{*, (k_1, \cdots, k_L)}(U)$, which we refer to as a {\it product} regularisation procedure.  \\
\begin{defn} Let $U=U_1\times \cdots \times U_L$ with $U_i, i=1, \cdots, L$ open subsets in $\R^n$ and let $(k_1, \cdots, k_L)$ be a multiindex of non negative integers. \\ 
Given a  product regularisation procedure $${\cal R}=\hat\otimes_{i=1}^L {\cal
  R}_i:\sigma=\otimes_{i=1}^k\sigma_i\mapsto \sigma(z)= \otimes_{i=1}^k
\sigma_i(z)$$ on $\hat\otimes_{i=0}^k  CS(U_i)$ of continuous regularisations ${\cal R}_i, i=1,\cdots, L$, the continuous  maps 
\begin{eqnarray*} 
CS^{\alpha_i}(U_i)&\to & \Ci(U_i,{\rm Hol}( \C- P_i))\\
\sigma_i&\mapsto&\left(x_i\mapsto  \int_{T_{x_i}^*U_i}\, {\cal R}_i(\sigma_i)(z)(x_i, \xi_i)\, d\xi_i\right),\quad i=1, \cdots, L\\
\end{eqnarray*}
induce  a uniquely defined map: 
\begin{eqnarray*} 
\hat\otimes_{i=0}^k  CS^{\alpha_i}(U_i)&\to & \Ci(U, {\rm Hol}(\C-\cup_{i=1}^k P_i))\\
\sigma &\mapsto&\left(x\mapsto  \altcutoffint_{T_{x}^*U}{\cal R}(\sigma)(z)(x, \xi)\, d\xi_1\cdots d\xi_L\right).\\
\end{eqnarray*}
Similarly the continuous maps 
\begin{eqnarray*} 
CS^{\alpha_i}(U_i)&\to & \Ci(U_i, \C)\\
\sigma_i&\mapsto&\left(x_i\mapsto  \cutoffint_{T_{x_i}^*U_i}^{{\cal R}_i}\sigma_i(x_i, \xi_i)\, d\xi_i\right),\quad i=1, \cdots, L\\
\end{eqnarray*}
induce a uniquely defined map:
\begin{eqnarray*} 
\hat\otimes_{i=0}^k  CS^{\alpha_i}(U_i)&\to & \Ci(U, \C)\\
\sigma &\mapsto&\left(x\mapsto  \altcutoffint_{T_{x}^*U}^{{\cal R}}\sigma(x, \xi)\, d\xi_i\right)\\
\end{eqnarray*}
which induces a linear map on $\hat\otimes_{i=0}^k  CS(U_i)$ called the multiple regularised integral associated with  the product regularisation ${\cal R}$. 
\end{defn}
The Wodzicki residue density ${\rm res}_{x_i}$  on $CS(U_i)$ similary give rise  by continuity to $\widetilde{\rm res}_{x, k}$ on $\hat\otimes_{i=1}^kCS(U_i)$ in such a way that for any $x=(x_1, \cdots, x_L)\in U_1\times\cdots\times U_L$:
$$\widetilde{\rm res}_{x,k}(\otimes\sigma_i(x_i,\cdot))= \prod_{i=1}^ k{\rm res}_{x_i}(\sigma_i(x_i, \cdot)).$$
\begin{thm}\label{thm:altiteratedreg}
Let $U=U_1\times \cdots \times U_L$ with $U_i, i=1, \cdots, L$ open subsets in $\R^n$ and let $\sigma\in \hat\otimes_{i=1}^k CS(U_i).$\\ 
Given a  product regularisation procedure $${\cal R}=\hat\otimes_{i=1}^L {\cal
  R}_i:\otimes_{i=1}^L\sigma_i\mapsto\otimes_{i=1}^L\sigma_i(z) $$ on $CS_w(U)$ of continuous regularisations ${\cal R}_i, i=1,\cdots, L$ such that ${\cal R}_i(\sigma)(z)$ has order $\alpha_i(z)$,
the map $$z\mapsto  \altcutoffint_{T_{x}^*U}{\cal R}(\sigma)(z)(x, \xi)\, d\xi_1\cdots d\xi_L$$ is meromorphic with poles at most of order $L$ and: 
$${\rm Res}^{L}_{z=0}\altcutoffint_{T_{x}^*U}{\cal R}(\sigma)(z)(x, \xi)\, d\xi_1\cdots d\xi_L= \frac{(-1)^L}{\prod_{i=1}^L\alpha_i^\prime(0)} \widetilde{\rm res}_{x, L}(\sigma).$$
In particular, when $\alpha_i^\prime(0)=\alpha^\prime(0)$ is constant this yields
 $${\rm Res}^{L}_{z=0}\altcutoffint_{T_{x}^*U}{\cal R}(\sigma)(z)(x, \xi)\, d\xi_1\cdots d\xi_L= \frac{(-1)^L}{\left(\alpha^\prime(0)\right)^L} \widetilde{\rm res}_{x, L}(\sigma).$$
\end{thm}
{\bf Proof:} By a continuity argument, this follows from the fact that this same relation holds on products $\sigma=\otimes_{i=1}^L \sigma_i$: 
\begin{eqnarray*}
{\rm Res}^{L}_{z=0}\altcutoffint_{T_{x}^*U}\prod_{i=1}^L{\cal R}_i(\sigma_i)(z)(x_i, \xi_i)\, d\xi_1\cdots d\xi_L&=& 
\prod_{i=1}^L{\rm Res}_{z=0}\cutoffint_{T_{x_i}^*U_i}{\cal R}_i(\sigma_i)(z)(x_i, \xi_i)\, d\xi_i\\
&=&\prod_{i=1}^L\frac{-1}{\alpha_i^\prime(0)} {\rm res}_{x_i}(\sigma_i)\\
&=&  \frac{(-1)^L}{\prod_{i=1}^L\alpha_i^\prime(0)} \widetilde{{\rm res}}_{x, L}(\sigma).
\end{eqnarray*}
\endsquare\\ \\
On the grounds of this theorem, taking finite parts we set:
\begin{defn}Given a  product regularisation  ${\cal R}=\hat \otimes_{i=1}^L {\cal R}_i$ on $ CS_w(U)$,    for any $\sigma\in CS_w(U)$ we call
$$\altcutoffint_{T_x^* U}^{\cal R}\sigma(x, \xi)\,::= {\rm fp}_{z=0}
\altcutoffint_{T_x^* U}  \sigma(z)(x, \xi)\, d\xi$$
with ${\cal R}:\sigma \mapsto \sigma(z)$,
the ${\cal R}$-regularised iterated integral of $\sigma$. 
\end{defn}
\begin{rk} With these notations we have: 
$$\altcutoffint_{T_{x_i}^* U}^{\cal R}d\xi \, \otimes_{i=1}^L\sigma_i(x_i,
\xi_i)=\prod_{i=1}^L \altcutoffint_{T_{x_i}^* U_i}^{{\cal R}_i}d\xi_i \,\sigma_i(x_i, \xi_i).$$
\end{rk}
\section{An alternative extension of regularised integrals to tensor products of classical symbols}
We now  give an alternative extension of regularised integrals to tensor products of classical symbols which we then compare with the one previously defined. 
For this purpose we consider a  map similar to the map  $\sigma\mapsto
\int_{\vert \xi \vert \leq R} \sigma(x, \xi) d\xi$ underlying the construction
of cut-off integrals. We will henceforth work under the assumption  $U_1=
\cdots= U_k=U$ an open subset of $\R^n$. 
\subsection{ Rota-Baxter relations }
\begin{prop}\label{prop:RotaBaxter}
\begin{enumerate}
\item The map $\sigma\mapsto P(\sigma)$ defined by 
$$P(\sigma)(x, \eta):= \int_{\vert \xi \vert \leq \vert \eta\vert}\sigma(x,\xi) \,d\xi$$
maps $CS^{*, k-1}(U)$ to $CS^{*, k}(U)$. Given $\sigma\in CS^{*,k-1}(U)$, $P(\sigma)=
C+\tau$ for some constant $C$ and with $\tau\in CS^{\alpha+n, k}$. In
particular, when $\alpha\in \R$, it has
order ${\rm max}(0, \alpha +n)$.\\ \\
\item For any $\sigma \in CS^{*, k-1}(U)$
\begin{equation}\label{eq:Pres}
P(\sigma)(x,\eta)- \frac{{\rm res}_{x, k-1}(\sigma)}{k} \, \log^{k}\vert \eta\vert \quad \in CS^{*, k-1}(U)
\end{equation}
 so that if $\sigma$ has vanishing residue of order $k-1$ then $P(\sigma)$ also lies in $CS^{*, k-1}(U)$.\\ \\
\item $P$ obeys the following Rota-Baxter relation \cite{EGK}:
\begin{equation}\label{eq:RotaBaxter}
P(\sigma)\,P( \tau)=  P(\sigma\, P(\tau))+ P(\tau \, P(\sigma)).
\end{equation}
\end{enumerate}
\end{prop}
{\bf Proof:} Replacing $R$ by $\vert \eta\vert $
in the asymptotic expansion (\ref{eq:cutoffasympt}) yields: 
\begin{eqnarray}\label{eq:Pasympt}
P(\sigma)(x,\eta)&\sim &C_x(\sigma)+ \sum_{j=0,\alpha-j+n\neq 0}^\infty \sum_{l=0}^{k -1}P_l(\sigma_{\alpha-j, l})(\log \vert \eta\vert)  \, \vert \eta\vert^{\alpha-j+n}\nonumber\\
&+&
\sum_{l=0}^{k-1} \frac{{\rm res}_{x,l}(\sigma)}{l+1}\log^{l+1}  \vert \eta\vert
\end{eqnarray} where $P_l(\sigma_{\alpha-j, l})(X)$ is a polynomial of degree $l$ with coefficients depending on $\sigma_{\alpha-j, l}$ and where  $C_x(\sigma)$ is the constant term corresponding to the finite part. \\
$P(\sigma)$ is therefore the sum of a symbol of order zero (the constant
$C_x(\sigma)$) and a symbol $\tau$  of order $\alpha+n$ so that when $\alpha\in \R$, its
order is ${\rm max}(0, \alpha +n)$. Furthermore, it lies in $CS^{\alpha, k}(U)$ and  the coefficient of 
$\log^{k}\vert \eta\vert$ is $\frac{{\rm res}_{x,k-1}(\sigma)}{k}.$\\
The Rota-Baxter relation then follows from:
\begin{eqnarray*}
P(\sigma)(\eta)\,P( \tau)(\eta)&=& \int_{\vert \xi \vert \leq \vert \eta\vert}\sigma(\xi) \,d\xi\, \int_{\vert \xi \vert \leq \vert \eta\vert}\tau(\xi) \,d\xi \\
&=& \int_{\vert \xi \vert \leq \vert \eta\vert}\sigma(\xi) \,d\xi\, \int_{\vert \tilde \xi \vert \leq \vert \xi\vert}\tau(\tilde \xi) \,d\, \tilde \xi
+ \int_{\vert \xi \vert \leq \vert \eta\vert}\tau(\xi) \,d\xi\, \int_{\vert \tilde \xi \vert \leq \vert \xi\vert}\sigma(\tilde \xi) \,d\, \tilde \xi
 \\
&=& P(\sigma\, P(\tau))(\eta)+ P(\tau \, P(\sigma))(\eta).\\
\end{eqnarray*}
\endsquare \\ \\
Let ${\cal C}_k:=  \hat\otimes_{i=1}^{k+1}CS^{*, *}(U_i) $ be the space of $k$-chains built from the $CS^{*, *}(U_i)$'s.  Using the   
 Rota-Baxter map  we define a map  $$P_\bullet:{\cal C}_{\bullet+1} \to {\cal C}_{\bullet }$$
by 
\begin{eqnarray*}
P_k: \hat\otimes_{i=1}^{k +1}CS^{*, *}(U_i)&\to &  \hat\otimes_{i=1}^{k} CS^{*, *}(U_i)\\
 P_k(\sigma) (\xi_1, \cdots,  \xi_{k})&:=& P\left(\sigma (\xi_1, \cdots,  \xi_{k},\cdot)\right)(\xi_{k}).\\
\end{eqnarray*} 
In particular we have:
$$P_k\left( \sigma_1\otimes \cdots \otimes \sigma_{k+1}\right)(\xi_1, \cdots, \xi_k)= \sigma_1(\xi_1)\cdots \sigma_k(\xi_k) \, P(\sigma_{k+1})(\xi_k).$$
\begin{thm}\label{thm:iteratedRotaBaxter} Let $U$ be an open subset of $\R^n$. For any integer  $k>1,$ 
\begin{enumerate}
\item the composition $P_1\circ \cdots \circ P_{k-1}$ maps  $\hat\otimes_{i=1}^k CS^{\alpha_i}(U)$ to $CS^{*, \, k-1}(U).$\\ 
For $ \sigma_i \in CS(U)$, 
\begin{equation} \label{eq:ProdP} 
 P_{1}\circ P_{2}\circ \cdots \circ P_{k-1}( \sigma_1\otimes \cdots\otimes \sigma_k)= P\left(\cdots P(\sigma_k)\sigma_{k-1}\cdots )\sigma_{2}\right)\, \sigma_1
\end{equation}
is a finite sum of log-polyhomogeneous symbols of order given by the partial
sum $\alpha_{1}+\alpha_{2}+\cdots
+\alpha_{j}+ (j-1)n$  with $
 j=1, \cdots, k$. \\
In particular,  when $\alpha_1, \cdots, \alpha_k\in \R$, then $P_{1}\circ P_{2}\circ
\cdots \circ P_{k-1}(\sigma)$ has order given by 
\begin{eqnarray*}
&{}&o\left(P_{1}\circ P_{2}\circ \cdots \circ P_{k-1}(\sigma)\right)\\
&=&{\rm max} 
\left( 0, \cdots, {\rm   max}(0, {\rm max} (0,
  \alpha_k+n)+\alpha_{k-1}+n),\cdots)+ \alpha_2+n\right) +\alpha_1.
\end{eqnarray*}

\item Furthermore,
\begin{equation} \label{eq:ProdPres}
P_1\circ \cdots \circ P_{k-1}(\sigma_1\otimes \cdots \otimes \sigma_k)(\xi_1)-
\frac{{\rm res}_{x}(\sigma_k)}{(k-1)!} \, \log^{k-1}\vert \xi_1 \vert \quad \in CS^{*, k-2}(U).
\end{equation}
\item The following shuffle  (or iterated Rota-Baxter) relations hold:
\begin{eqnarray}\label{eq:iteratedRotaBaxter}
\prod_{i=1}^k P(\sigma_i)&=&\sum_{\tau\in \Sigma_k}  
P\circ P_1\circ \cdots \circ P_{k-1}(\sigma_{\tau(1)}\otimes \cdots \otimes \sigma_{\tau(k)})\nonumber \\
&=&  \sum_{\tau\in \Sigma_k}  P\left(P\left(\cdots P(\sigma_{\tau(k)})\sigma_{\tau(k-1)}\cdots )\sigma_{\tau(2)}\right) \sigma_{\tau(1)}\right).
\end{eqnarray}
\end{enumerate}
\end{thm} 
\begin{rk}
For $k=2$ equation (\ref{eq:iteratedRotaBaxter}) yields    back equation (\ref{eq:RotaBaxter}).
\end{rk}
{\bf Proof:}
\begin{enumerate}
\item  By a continuity argument, it suffices to show that $ P_{1}\circ
  P_{2}\circ \cdots \circ P_{k-1} (\sigma)\in CS^{*, k-1}(U)$ for any
  $\sigma= \sigma_1\otimes\cdots\otimes \sigma_k$.  This  follows from the
  first point in Proposition \ref{prop:RotaBaxter}   by
  induction on $k$. Indeed, appplying it to   $k=2$, we first check that 
  $P_1(\sigma_1)\in CS^{*, 1}(U)$; then assuming that the statement holds for
  $k$ we can apply Proposition \ref{prop:RotaBaxter}  to $ P_{2}\circ
  P_{3}\circ \cdots \circ P_{k} ( \sigma_2\otimes\cdots\otimes \sigma_{k+1})\in
  CS^{*, k-1 }(U) $ from which we  infer that 
\begin{eqnarray*}
&{}& P_{1}\circ
  P_{2}\circ \cdots \circ P_{k} ( \sigma_1\otimes \sigma_2\otimes\cdots\otimes
  \sigma_{k+1})\\
&=& P\left(P_{2}\circ
  P_{3}\circ \cdots \circ P_{k}\left(\sigma_1\otimes
  \sigma_2\otimes\cdots\sigma_{k+1}\right)\right)\in CS^{*,k}(U). 
\end{eqnarray*}
This formula combined with Proposition \ref{prop:RotaBaxter} also yields in a
similar manner that  $P_1\circ P_{2}\circ
  P_{3}\circ \cdots \circ P_{k-1} (\sigma_1\otimes\cdots\otimes \sigma_k)$ is a finite sum of log-polyhomogeneous symbols of order $\alpha_{1}+\cdots
+\alpha_{j}+ (j-1)n$ with $ j=1, \cdots, k$. From there we easily   derive  the formula for  degree
of   $P_1\circ P_{2}\circ
  P_{3}\circ \cdots \circ P_{k-1} (\sigma_1\otimes\cdots\otimes \sigma_k)$ when the $\alpha_i$'s are
real.
\item Similarly, an induction  using   equation (\ref{eq:Pres})  implies  equation (\ref{eq:ProdPres}). 
\item
Equation (\ref{eq:iteratedRotaBaxter}) follows from equation (\ref{eq:RotaBaxter}) in a similar manner. 
\end{enumerate}
\endsquare
\subsection{ Iterated cut-off integrals of classical  symbols}
By  the results of the previous paragraph, the operator $P_{1}\circ \cdots
\circ P_{k-1}$ sends $\hat \otimes_{i=1}^k CS(U)$ to
$CS^{*, k-1}(U)$, a space  on which we
can apply cut-off regularisation described in section 2.
\begin{defn} Let  $U\subset \R^n$ be an  open subset. For $\sigma \in \hat
  \otimes_{i=1}^k CS(U)$  and given a point $x\in U$ we set
\begin{eqnarray*}
   \cutoffint_{T_x^*U} \sigma(\xi) d\xi&:=&\sum_{\tau\in \Sigma_k}
   \cutoffint_{T_{x}^* U}d\xi  P_{1}\circ \cdots \circ P_{k-1}(\sigma\circ \tau)(\xi) \\
&=&\sum_{\tau\in \Sigma_k}  \cutoffint_{T_{x_1}^* U_1}d\xi_{1}\int_{\vert \xi_{2}\vert \leq\vert \xi_{1}\vert} d\xi_{2}\cdots \int_{\vert \xi_k\vert \leq\vert \xi_{k-1}\vert} d\xi_{k}\, \sigma(\xi_{\tau(1)}, \cdots, \xi_{\tau(k)}) .
\end{eqnarray*}
\end{defn}
\begin{lem}  Let  $U\subset \R^n$ be an  open subset. For $\sigma_1, \cdots,
  \sigma_k \in CS(U)$ such all the (left) partial sums of the orders $\alpha_1+ \alpha_{2}+\cdots +\alpha_{j}$, $j=1,
  \cdots, k$   are  non integer valued, then
$$\prod_{i=1}^k \cutoffint_{T_x^*U_i} \sigma_i(x_i, \xi_i)\, d\xi_i= {\rm fp}_{R\to \infty}\prod_{i=1}^k  \int_{\vert \xi_i\vert \leq R} \sigma_i(x_i, \xi_i)\, d\xi_i.$$ 
\end{lem}
{\bf Proof:} We need to show that 
$$\prod_{i=1}^k {\rm fp}_{R_i\to \infty} \int_{\vert\xi_i\leq\vert R_i} \sigma_i(x_i, \xi_i)\, d\xi_i= {\rm fp}_{R\to \infty}\prod_{i=1}^k  \int_{\vert \xi_i\vert \leq R} \sigma_i(x_i, \xi_i)\, d\xi_i.$$ 
 For each $i\in \{1, \cdots, k\}$ we have the following asymptotic expansion (see equation (\ref{eq:cutoffasympt})):
\begin{eqnarray*}
\int_{\vert \xi_i\vert \leq R_i} \sigma_i(x_i,\xi_i) d\xi_i&\sim_{ R_i\to \infty}&C_x(\sigma_i)+ \sum_{m=0,\alpha_i-m+n\neq 0}^\infty \sum_{p=0}^{k_i} P_p(\sigma_{\alpha_i-m, p})(\log R_i) R_i^{\alpha_i-m+n}\\
&+&
\sum_{p=0}^{k_i} \frac{{\rm res}_{p,x_i}(\sigma_i)}{p+1}\log^{p+1}  R_i.
\end{eqnarray*}
Multiplying these asymptotic expansions and setting $R_i=R$ can give rise to new finite parts other than $\prod_{i=1}^k {\rm fp}_{R_i\to \infty} \int_{\vert\xi_i\vert \leq R_i} \sigma_i(x_i, \xi_i)\, d\xi_i= \prod_{i=1}^kC_x(\sigma_i).$ Indeed, when setting $R_i=R_j=R$,  positive powers of $ R_i$ arising from the asymptotic expansion of $ \int_{\vert\xi_i\vert\leq R_i} \sigma_i(x_i, \xi_i)\, d\xi_i$ 
 might compensate negative powers of $R_j$ arising from the asymptotic
 expansion of $ \int_{\vert\xi_j\vert\leq  R_j} \sigma_i(x_j, \xi_j)\, d\xi_j$
 thus leading to a new constant term. But since such powers arise in the form
 $R^{\alpha_1+\alpha_{2}+\cdots +\alpha_{j}-m+j\, n} $ such a compensation
 can only happen if $\alpha_1+\alpha_{2}+\cdots +\alpha_{j}$ takes integer values.   One therefore avoids such compensations assuming that
non of all the  (left) partial sums of the orders $ \alpha_1+\alpha_{2}+\cdots +\alpha_{j}$ are non  integers.\endsquare\\ \\
We deduce from the definition and the above lemma that cut-off regularisation "commutes" with products
of symbols in certain special cases: the cut-off iterated  integral of a
product of symbols coincides with the product of the cut-off integrals of the
symbols provided these have orders whose (left)   partial sums are non integer valued. 
\begin{prop}\label{prop:cutoffprod}
Let $\sigma_i\in CS^{\alpha_i}(U)$, $i=1, \cdots, k$ such that all the (left)
partial sums of the orders  $ \alpha_1+\alpha_{2}+\cdots +\alpha_{j}, j=1, \cdots, k$  are non integer valued. Then
\begin{equation}\label{eq:intprod}
  \cutoffint_{T_x^*U}\prod_{i=1}^k \sigma_i(x,\xi_i)\, d\xi_i= \prod_{i=1}^k \cutoffint_{T_{x_i}U_i}\sigma_i(x, \xi_i) \, d\xi_i.
\end{equation}
\end{prop}
{\bf Proof:}  From the above lemma it follows that 
\begin{eqnarray*}
&{}&\prod_{i=1}^k \cutoffint_{T_{x}U}\sigma_i(x, \xi_i) \, d\xi_i\\
&=&  {\rm fp}_{R\to \infty}\prod_{i=1}^k  \int_{\vert \xi_i\vert \leq R} \sigma_i(x, \xi_i)\, d\xi_i\\
&=&   {\rm fp}_{R\to \infty}\sum_{\tau \in \Sigma_k}  \int_{\vert \xi_1\vert \leq R} d\xi_{1} \int_{\vert \xi_{2}\vert\leq \vert\xi_1\vert}\cdots \int_{\vert \xi_{k}\vert \leq \vert \xi_{L-1}\vert} d\xi_k\prod_{i=1}^k  \sigma_{\tau(i)} \left(x, \xi_{\tau(i)}\right)\\
&=& \cutoffint_{T_x^*U}\prod_{i=1}^k \sigma_i(x,\xi_i)\, d\xi_i.
\end{eqnarray*}
\endsquare \\ \\
\begin{thm}\label{thm:regiteratedRotaBaxter}
Let $\sigma_i\in CS^{\alpha_i}(U)$, $i=1, \cdots, k$ be such that all the (left)
partial sums   $\alpha_1+\alpha_{2}+\cdots +\alpha_{j},\quad j=1, \cdots, k$ are non integer valued. Then the following shuffle relations hold:
\begin{eqnarray}\label{eq:regiteratedRotaBaxter}
&{}& \prod_{i=1}^k \cutoffint_{T_x^*U} d\xi_i\, \sigma_i\nonumber\\
 &=& \sum_{\tau\in \Sigma_k}  \cutoffint_{T_{x}^*U} \, d\xi\, P_1\circ \cdots\circ P_{k-1} \left(\sigma_{\tau(1)} \otimes\cdots\otimes \sigma_{\tau(k)}\right)(\xi)\,d\xi\\
&=& \sum_{\tau\in \Sigma_k}  \cutoffint_{T_{x}^* U}d\xi_{1}\int_{\vert \xi_{2}\vert \leq\vert \xi_{1}\vert} d\xi_{2}\cdots \int_{\vert \xi_{k-1}\vert \leq\vert \xi_{k}\vert} d\xi_{k-1}\, \sigma_{\tau(k)}(\xi_{k}) \cdots \sigma_{\tau(1)}(\xi_1).\nonumber
\end{eqnarray}
\end{thm}
{\bf Proof:} Recall that $P(\sigma_i)(\eta_i)=\int_{\vert \xi\vert \leq \vert
  \eta_i\vert} \sigma_i(\xi) \, d\xi.$ Applying  equation
(\ref{eq:iteratedRotaBaxter}) to $\eta_i= R$ for $i=1, \cdots k$  and then taking the finite part when $R\to \infty$  yields the result:
\begin{eqnarray*}
\prod_{i=1}^k \cutoffint_{T_{x}^*U} \sigma_i &=&
 \prod_{i=1}^k  {\rm fp}_{R\to \infty}\,\int_{B_x(0, R)}  \sigma_i \\
&=&  {\rm fp}_{R\to \infty}\left( \sum_{\tau\in \Sigma_k} \int_{B_x(0, R)}
P_{1}\circ \cdots \circ P_{k-1}(\sigma_{\tau(1)}\otimes \cdots \otimes \sigma_{\tau(k)}) \right)\\
&=&  \sum_{\tau\in \Sigma_k}  \cutoffint_{T_{x}^*U} \,  P\left(\cdots P(\sigma_{\tau(k)})\sigma_{\tau(k-1)}\cdots )\sigma_{\tau(2)}\right) \sigma_{\tau(1)}.
\end{eqnarray*}
The above lemma then yields the result under the assumption that all partial orders are non integer. \endsquare
\subsection{Iterated integrals of holomorphic families of classical  symbols}
When the symbols have integer order, neither  does the iterated cut-off integral of the tensor  product of the symbols  coincide with the product of their cut-off integrals (see  equation (\ref{eq:intprod})), 
 nor do  the shuffle relations (\ref{eq:regiteratedRotaBaxter}) hold for
 cut-off integrals. However  holomorphic perturbation of these symbols will
 have holomorphic orders, the (left) partial sums of which will be non integer
 outside a discrete set and both equation (\ref{eq:intprod}) and the shuffle
 relations (\ref{eq:regiteratedRotaBaxter}) hold for these perturbed
 symbols.\\ \\
\begin{prop}\label{lem:meroint}
Let $
 {\cal R}: \sigma\mapsto \sigma(z)$ be a  holomorphic  regularisation procedure
 on $ CS^{*,*}(U)$ such that $\sigma(z)$ has order $\alpha(z)=q\, z+\alpha(0)$
 with $q\neq 0$. \\
 For any $\sigma_i\in CS^{*, k_i}(U), i=1, 2$, with $\sigma_i(z)$ of order
 $\alpha_i(z)= q\, z+ \alpha_i(0)$
\begin{enumerate}
\item the map
$$z\mapsto \cutoffint_{T_x^*U} P(\sigma_2(z))(\xi)\, \sigma_1(z)(\xi)\, d\xi$$
is meromorphic with at most poles of order $k_1+k_2+2$ in the discrete set 
 $$P_2:= q^{-1}\, \left( \Z- \alpha_1(0)\right)\, \cup \, (2\, q)^{-1} \left( \Z-
  \alpha_1(0)-\alpha_2(0)\right).$$
\item
We have the following identity of meromorphic functions: 
\begin{eqnarray}\label{eq:meroshuffle2}
&{}& \cutoffint_{T_{x}^*U} d\xi_1\, \sigma_1(z)\,\cutoffint_{T_{x}^*U} d\xi_2\, \sigma_2(z) \\
 &=&  \cutoffint_{T_{x}^*U} \, P
\left(\sigma_{1}(z)\right)(\xi)\, \sigma_{2}(z)(\xi) \, d\xi+ 
 \cutoffint_{T_{x}^*U} \,  P
\left(\sigma_{2}(z)\right)(\xi)\, \sigma_{1}(z)(\xi) \, d\xi.\nonumber  
\end{eqnarray}
\end{enumerate} 
\end{prop}
{\bf Proof:}
\begin{enumerate}
\item  We first observe that $P(\sigma_2(z))\,
\sigma_1(z)$  is the sum of 
a  symbol $\tau_1(z)\in CS^{\alpha_1(z) , k_1}(U)$  proportional to $\sigma_1(z)$  and a
 symbol $\tau_2(z)\,\sigma_1(z) \in CS^{\alpha_1(z)+\alpha_2(z)+n,k_1+k_2+ 1}(U)$ 
with $\tau_2(z)\in CS^{\alpha_2(z)+n,k_2+ 1}(U)$
 (see
 Proposition \ref{prop:RotaBaxter}). By Theorem \ref{thm:KV}   and using  the
 linearity of the cut-off
integral, we find that  the cut-off integral $$\cutoffint_{T_x^*U}
P(\sigma_2(z))(\xi)\, \sigma_1(z)(\xi)\, d\xi= \cutoffint_{T_x^*U}\tau_1(z)(x,
\xi) \, d\xi+ \cutoffint_{T_x^*U}\tau_2(z)(
\xi) \, \sigma_1(z)(\xi)\,d\xi$$  is meromorphic  with  poles of order at
most $k_1+k_2$ at points in $P_2$ defined as in the proposition since $\alpha_1(z)= q\,
z+\alpha_1(0)$ and  $\alpha_1(z)+\alpha_2(z)+n =
2q\, z+\alpha_1(0)+\alpha_2(0)+n$. 
\item  Equation (\ref{eq:meroshuffle2}) then follows from  applying
  (\ref{eq:regiteratedRotaBaxter}) to  $\sigma_i:= \sigma_i(z)$  (with $k=2$)
  outside the discrete set of poles.
\end{enumerate}
\endsquare\\ \\
This generalises to the tensor product of  $k$ symbols.
\begin{thm}\label{thm:holprodreg}
 Let  $U$ be an open subset of $R^n$ and let $
 {\cal R}$ be a holomorphic regularisation procedure
 $\sigma\mapsto \sigma(z)$ on $ CS(U)$ such that $\sigma(z)$ has order
 $\alpha(z)=q\, z+\alpha(0)$ with $q\neq 0$.\\
 For any
 $\sigma_i\in CS(U)$ with $\sigma_i(z)$ of order 
$\alpha_i(z)=q\, z+\alpha_i(0)$
\begin{enumerate}
\item the map $z\mapsto \cutoffint_{T_{x}^*U} \, d\xi\,P_1\circ \cdots \circ
 P_{k-1}
\left(\sigma_1(z)\otimes \cdots\otimes\sigma_k(z)\right)(\xi)$ is
meromorphic with  poles of  order at most $k$ in
$${\cal P}_k:= \bigcup_{j=1}^k  (j\,q)^{-1}\,\left(\Z-\alpha_1(0)-\alpha_{2}(0)-\cdots -\alpha_{j}(0)\right).$$ 
\item The map 
$$z\mapsto \cutoffint_{T_x^*U}\otimes_{i=1}^k \sigma_i(z)\, d\xi$$ is meromorphic  with poles  of order at most  $k$  and we have the following equality of  meromorphic functions: 
\begin{eqnarray}\label{eq:meroshuffle}
&{}& \prod_{i=1}^k \cutoffint_{T_{x}^*U} d\xi_i\, \sigma_i(z)\nonumber\\
 &=& \sum_{\tau\in \Sigma_k}  \cutoffint_{T_{x}^*U} \, P_1\circ \cdots \circ P_k
\left(\sigma_{\tau(1)}(z)\otimes \cdots\otimes\sigma_{\tau(k)}(z)\right)(\xi) \, d\xi,
\end{eqnarray}
where $\Sigma_k$  denotes the group of permutations on $k$ elements.
\end{enumerate}
\end{thm}
{\bf Proof:} 
Statements 1 and 2 in the theorem  follow by induction on k from statements 1 and 2 of
Proposition \ref{lem:meroint}. Indeed, Proposition \ref{lem:meroint} with
$k_1=k_2=0$ yields   the theorem for $k=1$. Replacing   $\sigma_2$  in
Proposition 
\ref{lem:meroint} by  $P_2\circ \cdots \circ P_k (\sigma_2\otimes\cdots
\sigma_{k+1})\in CS^{*, k-1}(U)$ (so that $k_2= k-1$ here) then yields   the induction step $k\to k+1$ since 
\begin{eqnarray*}
&{}& P_{1}\circ
  P_{2}\circ \cdots \circ P_{k} ( \sigma_1(z)\otimes
  \sigma_2(z)\otimes\cdots\otimes \sigma_{k+1}(z))\\
&=& P\left(P_{2}\circ
  P_{3}\circ \cdots \circ P_{k}\left(\sigma_1(z)\otimes
  \sigma_2(z)\otimes\cdots\sigma_{k+1}(z)\right)\right). 
\end{eqnarray*}
\endsquare
\begin{cor}\label{cor:regshuffle}
Under the same assumptions and using  the same notations as in Theorem   \ref{thm:holprodreg},  we have the following equality of meromorphic maps:
\begin{eqnarray} \label{eq:prodreg}
\cutoffint_{T_x^*U}\otimes_{i=1}^k \sigma_i(z)\, d\xi &=& \altcutoffint_{T_x^*U}\otimes_{i=1}^k \sigma_i(z)\, d\xi\nonumber\\
&=& 
\prod_{i=1}^k \cutoffint_{T_x^*U_i}\sigma_i(z)(x_i, \xi_i)\, d\xi_i.
\end{eqnarray}
The highest order pole is given by:
$${\rm Res}_{z=0}^{k} \cutoffint_{T_x^*U}\otimes_{i=1}^k \sigma_i(z)\, d\xi=
\frac{(-1)^k}{\prod_{i=1}^k \alpha_i^\prime(0)} \widetilde{{\rm res}}_{x,
  k}(\otimes_{i=1}^k\sigma_i)= \prod_{i=1}^k\frac{-1}{ \alpha_i^\prime(0)}
{\rm res}_{x}(\sigma_i).$$
\end{cor}
{\bf Proof:} As a consequence of   the shuffle relations
(\ref{eq:meroshuffle}), we have the following equality of meromorphic functions 
\begin{eqnarray*} 
\cutoffint_{T_x^*U}\otimes_{i=1}^k \sigma_i(z)\, d\xi &=&  \sum_{\tau\in \Sigma_k}  \cutoffint_{T_{x}^*U} \, P_1\circ \cdots \circ P_k
\left(\sigma_{\tau(1)}(z)\otimes \cdots\otimes\sigma_{\tau(k)}(z)\right)(\xi) \, d\xi\\
&=&
\prod_{i=1}^k \cutoffint_{T_x^*U_i}\sigma_i(z)(x_i, \xi_i)\, d\xi_i.\\
\end{eqnarray*}
On the other hand, by the results of section 3 we have a further equality of
meromorphic functions:
$$ \altcutoffint_{T_x^*U}\otimes_{i=1}^k \sigma_i(z)\, d\xi=
\cutoffint_{T_x^*U}\otimes_{i=1}^k \sigma_i(z)\, d\xi, $$
which shows that the two regularised integrals $\cutoffint$ and
$\altcutoffint$ both  coincide on tensor products of holomorphic symbols with the product of the
regularised integral of each of the symbols. The Wodzicki residue formula
then follows from Theorem \ref{thm:altiteratedreg}.\endsquare 
\subsection{Obstructions to shuffle relations for regularised integrals of general classical symbols}
The finite part of a product of meromorphic functions with poles generally
does not coincide with the product of the finite parts. As a result,  when the  symbols have non vanishing  residues, 
taking finite parts of  the above shuffle relations on the level of  meromorphic functions
does not yield the expected shuffle equations for the corresponding finite parts.  However, in that case a  renormalisation procedure familiar to physicists   provides the obstruction in terms of counterterms arising in the renormalisation.\\
Let ${\cal M}(\C)$ denote the algebra of meromorphic functions on $\C$, and let ${\cal M}^k(\C)$ denote the space of meromorphic functions on $\C$ with
poles of order at most $k$  at $z=0$. Clearly, if $f_1, \cdots, f_k \in {\cal
  M}^1(\C)$ then $\prod_{i=1}^k f_i \in {\cal M}^k(\C)$. Let as before ${\fp}_{z=0} f= \lim_{z\to 0} z\, f(z)$ denote the finite part at $z=0$ of a function $f\in {\cal M}^1(\C)$. Then, in general 
$$\prod_{i=1}^k {\fp}_{z=0} f_i(z)\neq {\fp}_{z=0} \prod_{i=1}^k  f_i(z).$$ 
A renormalisation procedure taken from  physics provides a recursive procedure
to compute the obstruction to the equality; when  the products  $
\prod_{i=1}^k  f_i(z)$  arise from applying dimensional regularisation to
Feynman type functions in the language of Etingof \cite{E},  this comes down
to applying the renormalisation procedure used by physicists for connected
Feynman graphs to a concatenation of disjoint one loop diagrams. \\ \\
The
underlying Hopf algebra (\cite{K2},\cite{CK}) in the situation considered here is
the symmetric algebra ${\cal H}:= \oplus_{k=0}^\infty\bigodot^k CS(U)$ \footnote{$\odot$ denotes the symmetrised tensor product.} built on the 
vector space $CS(U)$. It is in particular commutative and
cocommutative. Although very simple, this toy model is instructive.  The
(deconcatenation) coproduct on $\sigma=\sigma_1\odot
\cdots \odot \sigma_k$  reads:  $$\Delta \sigma=
\sigma\otimes 1+ 1\otimes \sigma +  \sum_{J\subsetneq \{1,\ldots
  ,k\},
  J\neq \phi}  \bigodot_{j\in J} \sigma_j  \odot   \bigodot_{i\notin J}
\sigma _i.$$
A regularisation procedure ${\cal R}: \sigma\mapsto \sigma(z)$ induces a map
$\phi: CS(U)\to   {\mathcal  M}^1(\C)$ defined by 
$$\phi(\sigma)(z)= \cutoffint_{T_x^*U} \sigma(z)(x, \xi)\, d\xi.$$
 
Our previous constructions show it extends to an algebra morphism 
\begin{eqnarray*}
\Phi: {\cal H} &\to& {\cal M}(\C)\\
\sigma=\sigma_1\otimes \cdots \otimes\sigma_k &\mapsto & \cutoffint_{T^* U\times
  \cdots \times T^* U }\sigma_1(z)(x, \xi_1)\cdots \sigma_k(z)(x, \xi_k) d\xi_1\cdots d\xi_k.
 \end{eqnarray*}
The Hopf structure on ${\cal H}$ provides a recursive procedure to get a
Birkhoff decomposition of the corresponding loop
 $\Phi(\sigma)$ for any $\sigma\in {\cal H}$  i.e. a factorisation of the form
$$\Phi(\sigma) = \Phi_-(\sigma)^{-1} \, \Phi_+(\sigma),$$
where $ \Phi_+(\sigma)$ is holomorphic at $0$. Namely, with Sweedler's
notations $\Delta x=x\otimes 1+ 1\otimes x+ \sum x^\prime \otimes x^{\prime \prime}$
$$\Phi_-(\sigma):= -T\left(\Phi(\sigma) + \sum \Phi_-(\sigma^\prime)
  \Phi(\sigma^{\prime\prime})\right),$$
$$\Phi_+(\sigma):=\Phi(\sigma) +\Phi_-(\sigma) +   \sum \Phi_-(\sigma^\prime)
\Phi(\sigma^{\prime \prime})       ,$$
where $T$ is the  projection on the pole part. 
This corresponds to  Bogolioubov's prescription by which one   first  "prepares"
\footnote{We borrow this expression and the notations that follow from
  \cite{CM} but we refer the reader to  Kreimer \cite{K2}, see also \cite{CK}
  for the Hopf algebra that underlies this renormalisation procedure.} the
symbol $\sigma$.
\\ \\
There is another way of describing this renormalisation procedure via a renormalisation operator $R$ on the space of  Laurent series $(z_1,\cdots, z_k)\mapsto f(z_1, \cdots, z_k)$ in several variables. 
For this,  instead of $$\Phi(\sigma): z\mapsto  \cutoffint_{T_x^* U\times
  \cdots \times T_x^* U }\sigma_1(z)(x, \xi_1)\cdots \sigma_k(z)(x, \xi_k)\,
d\xi_1\cdots d\xi_k,$$ let us  consider the map $$(z_1, \cdots, z_k)\mapsto  \cutoffint_{T_x^* U\times
  \cdots \times T_x^* U }\sigma_1(z_1)(x, \xi_1)\cdots \sigma_k(z_k)(x, \xi_k)
\, d\xi_1\cdots d\xi_k$$ which defines a Laurent series in  $(z_1, \cdots,
z_k)$;  setting $z_1=z_2=\cdots =z_k=z$ gives back the meromorphic function
$\Phi(\sigma)$. Given  a subset $J=\{i_1,\cdots, i_{\vert J\vert}\}\subsetneq
\{1, \cdots, k\}$,   setting $\bar J:=\{i_{\vert J\vert +1},\cdots, i_{\vert
  K\vert}\}$ to be its complement in $\{1, \cdots, k\}$,  from  such a Laurent
series $f$  we build the map $$f_J:(z, z_{i_{\vert J\vert+1}}, \cdots
z_{i_{\vert \overline{ J}\vert}})\mapsto f(z_1, \cdots, z_{i_{\vert K\vert}})_{\vert_{z_{i}=z, \forall i\in J}}.$$
When $f=\Phi(\sigma_1\otimes \cdots\otimes \sigma_k)= f_1\otimes\cdots\otimes  f_k$ with $f_i=\phi(\sigma_i)$ then 
$f_J(z, z_{i_{\vert J\vert +1}}, \cdots z_{i_{\vert \bar J\vert}})= \prod_{j\in J} f_j(z) \cdot \prod_{j\in \overline{ J}} f_j (z_j).$
Let us set 
$$\bar R   (f)(z) :=  f(z_1,\cdots, z_k)_{\vert_{z_i=z,  1\leq i\leq k}}+\sum_{\phi\neq J\subsetneq \{1,\ldots
  ,k\}} C\left(  f_J(z, z_{i_{\vert J\vert+1}}, \cdots, z_{i_{\vert \bar J\vert}}) \right)_{z_{i_j}=z}$$ which, in the case $f=\otimes_{i=1}^ k f_i$ considered above reads
$$\bar R   (\otimes_{i=1}^ k f_i)(z) :=  \prod_{i=1}^k f_i(z)+\sum_{\phi\neq J\subsetneq \{1,\ldots
  ,k\}} C( \otimes_{j\in J} f_j)(z) \,  \prod_{i\notin J} f_i(z).$$
The counterterm $C$ is defined inductively on $\vert J\vert$  by 
$$C\left(  f_J(z, z_{i_{\vert J\vert+1}}, \cdots, z_{i_{\vert \bar J\vert}}) \right):= -T \left( \bar R\big( f_J(z, z_{i_{\vert J\vert+1}}, \cdots, z_{i_{\vert \bar J\vert}}) \big)\right)       $$
where  $T$ is the projection onto the pole part of the Laurent series in $z$.\\
The renormalisation operator $R$ is then defined by
\begin{eqnarray*}
R(f)& :=& \bar R (f)+ C(f)\\
&=& (1-T)(f)+(1-T) (\sum_{\phi \neq J\subsetneq \{1,\ldots
  ,k\}}T \left( C( f_J )\right) ),\\
\end{eqnarray*}
which for $f=\otimes_{i=1}^k f_i$ reads: 
\begin{eqnarray*}
R(\otimes_{i=1}^k f_i)& :=& \bar R (\otimes_{i=1}^k f_i)+ C(\otimes_{i=1}^k f_i)\\
&=& (1-T)(\prod_{i=1}^k f_i)+(1-T) (\sum_{J\subsetneq \{1,\ldots
  ,k\},
  J\neq \phi}T \left( C(\otimes_{j\in J} f_j )\right)\,  \prod_{i\notin J} f_i .\\
\end{eqnarray*}
To illustrate this construction, let us take $k=2$ and compute
$R(f)$ with $f$  a Laurent series in $z_1, z_2$ in each variable $z_i$. There are only two  subsets $J\subset\{1,2\}$ to consider in the renormalisation procedure $J_1=\{1\}$ and $J_2=\{2\}$ and we  set $f_i:= f_{J_i}$ so that 
$$R(f)= (1-T) (f)- (1-T) \left(T (f_1) +  T (f_2)\right).$$ 
%
 Writing $$f(z_1, z_2)= \sum_{-I\leq i\leq 1; -J\leq j\leq 1} a_{i\, j}z_1^i\, z_2^j+  o({\rm sup}(\vert z_1\vert, \vert z_2\vert ) ),$$ 
where $I$, resp. $J$ is  the largest order of the poles at $0$ of $f_1$, resp.  $f_2$  respectively, we get
\begin{eqnarray*}
R(f)(z)&=& (1-T)\left(  \sum_{-I\leq i\leq 1; -J\leq j\leq 1} a_{i\, j}z^{i+j} +  o( z)\right)\\
&- &(1-T)  \left(\left(\sum_{ i>0}a_{i\, j}z^i\, z_2^j\right)_{\vert_{z_2=z}}  +   \left(\sum_{ j>0} a_{i\, j}z_1^i\, z^j\right)_{\vert_{z_1=z}}\right)\\
&=&    \sum_{0\leq i+j} a_{i\, j}z^{i+j} +  o( z)\\
&- & \left(\sum_{ i>0, i+j\geq 0}a_{i\, j}z^{i+j}  +   \sum_{ j>0, i+j\geq 0} a_{i\, j}z^{i+j}\right)\\
\\
&=& a_{0\, 0} +o(1).
\end{eqnarray*}
In particular, for two meromorphic functions $f_1$ and $f_2$ with simple poles:
$${\rm fp}_{z= 0} \left(R(f_1\otimes  f_2)\right)(z)= R(f_1\, f_2)(0)= {\rm fp}_{z=0}f_1(z) \,
{\rm fp}_{z=0} f_2(z).$$
More generally, an induction procedure yields:  
\begin{thm}\label{thm:regprod}
Let $(z_1, \cdots, z_k)\mapsto f(z_1, \cdots, z_k)$ have  a Laurent expansion  in each of the variables $z_i$.   The map $z\mapsto  R(f)(z)$ is holomorphic at $z=0$ and  its value at $z=0$ coincides with the constant term in the Laurent expansion in $(z_1, \cdots, z_k)$. \\
In particular, when $f=\otimes_{i=1}^k f_i$ where the functions  $f_i, i=1, \cdots, k$  are meromorphic  at $z=0$,
 then $R(f)(0)$ coincides with  the  product of the finite parts of  the $f_i$'s:
$${\rm fp}_{z\to 0} \left(R(f_1\otimes\cdots \otimes  f_k)(z)\right)=
R(f_1\otimes \cdots\otimes f_k)(0)=\prod_{i=1}^k  {\rm fp}_{z=0}f_i(z).$$
\end{thm}
{\bf Proof}: The operator $R$ yields  an algebra morphism on the algebra of Laurent series and takes values in meromorphic functions which are holomorphic at
$z=0$ \cite {CK}. As $f\mapsto R(f)(z)$  restricted to  ${\cal M}(\C)$ takes $f$ to a holomorphic function at $0$ with value $ R(f)(0)$ given by the finite part of $f$ at $z=0$, on a tensor product $f_1\otimes\cdots\otimes f_k\mapsto
R(f_1\otimes \cdots\otimes f_k)(0)$ picks up  the product of the finite
parts of the $f_i$'s at $z=0$. By a  closure argument, we conclude that the map $z\mapsto  R(f)(z)$ is holomorphic at $z=0$ on the whole algebra of Laurent series  and that  its value at $z=0$ coincides with the constant term in the Laurent expansion in $(z_1, \cdots, z_k)$. The second assertion is straightforward.
\endsquare \\
\begin{rk} As a consequence, if  instead of using one complex parameter $z$, we regularise each 
$\sigma_i$ by $\sigma_i\mapsto \sigma_i(z_i)$  using a different complex parameter $z_i$ we can avoid this renormalisation procedure: 
$${\rm fp}_{z_1,\cdots, z_k\to 0} \left(\otimes_{i=1}^k \cutoffint \sigma_i(z_i)\right)= {\rm fp}_{z_1,\cdots, z_k\to 0} (\otimes_{i=1}^k f_i)(z_1,\cdots,  z_k)=\prod_{i=1}^k  {\rm fp}_{z=0}f_i(z).$$
\end{rk}
Applying   the above theorem   to $f_i: z\mapsto \cutoffint_{T_{x_i}^*U_i}\sigma_i(z)$ we get the following description of the obstructions to shuffle relations for general classical symbols: 
\begin{cor}\label{cor:renshuffle}
Given a  regularisation procedure  ${\cal R}$ on $ CS(U)$  for any $i=1, \cdots, k$, for any $\sigma_i \in CS(U)$,
\begin{eqnarray*}
&{}&  \prod_{i=1}^k  \cutoffint_{ T_x^*U}^{\cal R} \sigma_i- \sum_{\tau\in
  \Sigma_k}\cutoffint_{T_{x_k}^* U}^{\cal R}  d\xi_k \, P_{1}\circ \cdots
\circ P_{k-1}\left( \sigma_\tau\right)\\
&=& {\rm fp}_{z=0} \cutoffint_{ T_x^*U\times \cdots T_x^*U }\left( f_1(z)\cdots f_k(z)- R(f_1\otimes\cdots\otimes f_k)(z)\right)\\
&=& \sum_{i_1+\cdots +i_k=0, (i_1, \cdots, i_k)\neq 0} a_{i_1}^1 \cdots a_{i_k}^k,
\end{eqnarray*}
where as before, $\sigma_{\tau}(i):= \sigma_{\tau(i)}$
 and
where the $a_i$'s correspond to the coefficients in the meromorphic expansion at $z=0$ of the cut-off integrals  $\cutoffint_{T_{x}^*U}\sigma_i(z)= \frac{a_{-1}^i}{z} + a_0^i + a_1^i \,z + o(z)$.\\
In particular, the shuffle relations therefore hold if all the $\sigma_i$'s have vanishing residue.\\
\end{cor}
{\bf Proof:} As in the proof of Corollary \ref{cor:regshuffle}
we have
 \begin{eqnarray*}
&{}& {\rm fp}_{z=0} \left[\prod_{i=1}^k \cutoffint_{T_{x}^*{U}}\sigma_i(z) \, d\xi_i\right]= \\
 &=& \sum_{\tau\in \Sigma_k}  \cutoffint_{T_{x}^*U}^{{\cal R}} \, d\xi_1 P\left(\cdots P(\sigma_{\tau(k)})\sigma_{\tau(k-1)}\cdots )\sigma_{\tau(2)}\right)(\xi_1) \sigma_{\tau(1)}(\xi_1)\nonumber\\
&=& \sum_{\tau\in \Sigma_k}  \cutoffint_{T_{x}^* U}^{{\cal R}}d\xi_{1}\int_{\vert \xi_{2}\vert \leq\vert \xi_{1}\vert} d\xi_{2}\cdots \int_{\vert \xi_{k}\vert \leq\vert \xi_{k-1}\vert} d\xi_{k}\, \sigma_{\tau(1)}(\xi_{1}) \cdots \sigma_{\tau(k)}(\xi_k).
\end{eqnarray*}
On the other hand, Theorem \ref{thm:regprod} applied to  $f_i: z\mapsto \cutoffint_{T_{}^*U}\sigma_i(z)$ yields 
\begin{eqnarray*}
&{}& {\rm fp}_{z=0}\left(\prod_{i=1}^k \cutoffint_{T_{x_i}^*U_i}\sigma_i(z) - R\left(\bigotimes_{i=1}^k \cutoffint_{T_{x}^*U}\sigma_i(z)\right)\right)\\
&=&{\rm fp}_{z=0}\left[\prod_{i=1}^k \cutoffint_{T_{x}^*U}\sigma_i(z)\right] - \prod_{i=1}^k \cutoffint_{T_{x}^*U}^{\cal R}\sigma_i(z)\\
&=&
 \sum_{i_1+\cdots +i_k=0, (i_1, \cdots, i_k)\neq 0} a_{i_1}^1 \cdots a_{i_k}^k\\
\end{eqnarray*} which in turn yields the result of the theorem. \endsquare
\section{Relation to multiple zeta functions}
We want to adapt the previous results to symbols of operators on the unit circle. But instead of using an atlas on $S^1$ and expressing the symbol of the operators in local charts (e.g. using stereographic projections), we  view $S^1$ as the Lie group $U(1)$  seen as the range of $(\R, +)$  under the group morphism:
\begin{eqnarray*}
\Phi: \R&\to & S^1\\
x &\mapsto & e^{i\, x}
\end{eqnarray*}
which has kernel $2\pi \Z\simeq \pi_1(S^1)$. This amounts to identifying $S^1$
with the quotient $\R/2\pi\Z$. In this picture, the additive  group structure
on $\R/2\pi\Z$ is identified with the multiplicative group structure on $S^1$:
$$\Phi(x+y+2\pi n)= \Phi(x+ 2\pi k)\Phi(y+2\pi l)\quad\forall k, l, n\in \Z,$$
an important fact for what follows. \\ 
\subsection{The symbol of invariant operators on the unit circle}

We then identify $S^1$ with $\R/2\pi\Z$ and note the group law additively. The
kernel $K(x,y)$ of an invariant operator $P$ depends only on the difference
$x-y$. It lifts to a $2\pi$-periodic function $\tilde K$ on $\R$. The Fourier
transform of $\tilde K$ is a linear combination of Dirac masses at the integers, and can
reasonably be taken as a symbol for the operator $P$. It defines then a
$S^1$-invariant distribution on the cotangent $T^*S^1$. The trace of $P$, when
it exists, will be given by the integral of the symbol on $T^*S^1$.\\ \\
\qquad
We will illustrate this principle on complex powers of the laplacian. The Laplacian  $$\Delta=-\partial_t^2$$ on $S^1$ has discrete spectrum
$\{n^2, n\in \Z\}$. The operator $\Delta^\prime:=\Delta_{\vert_{{\rm
      Ker}\Delta^\perp}}$ where ${\rm
      Ker}\Delta^\perp$ denotes the orthogonal space to the kernel,  has spectrum 
$\{n^2;   n\in \Z-\{0\}\}$ and its square root $\sqrt{\Delta^\prime}$ has spectrum
$$\{\vert n\vert,    n\in \Z-\{0\}\}$$ as a consequence of which its  zeta function  is given by:
\begin{eqnarray*}
\zeta_{\sqrt {\Delta^\prime}}(z)&:=&\sum_{n\in \Z-\{0\}}\vert n\vert ^{-z}\\
&=&  2\sum_{n=1}^\infty n^{-z}= 2\zeta(z)
\end{eqnarray*}
where $\zeta$ is the Riemann zeta function.\\
$\zeta_{\sqrt {\Delta^\prime}}(z)$ can also be seen as the canonical trace of the operator $\sqrt {\Delta^\prime}^{-z}$ so that: 
\begin{eqnarray*}
\zeta_{\sqrt {\Delta^\prime}}(z)&=& {\rm TR}\left(\sqrt {\Delta^\prime}^{-z}\right)\\
&=& \int_{T^*S^1} \sigma_z(x,\xi) dx\, d\xi
\end{eqnarray*}
where $\sigma_z$ is the symbol of $\sqrt {\Delta^\prime}$ (still to be defined). 
We use the Mellin transform to express $\sqrt {\Delta^\prime}^{-z}$ in terms
of the heat-kernel of $\Delta$ on $S^1$: 
$$\sqrt {\Delta^\prime}^{-z}=\frac{1}{\Gamma(\frac z2)} \int_0^\infty t^{\frac
  z2-1} e^{-t\Delta^\prime}\, dt. $$
We want to  compute its symbol. 
\begin{prop}\label{prop:symbolofdeltaz}
The symbol of $\sqrt {\Delta^\prime}^{-z}$ where $\Delta$ is the Laplacian on $S^1$ reads for $\xi\in \R$: 
$$\sigma_z(x,\xi)=  \sum_{k\in\Z-\{0\}} \vert k\vert^{-z}
\delta_{k}(\xi)$$
\end{prop}
{\bf Proof :} If $H_t(x, y)= h_t(x-y)$ denotes the heat-kernel of $\Delta$ on $S^1$ we have for every $f\in\Ci(S^1, \R)\cap{\rm Ker}\Delta^\perp$:   
$$\left(\sqrt {\Delta^\prime}\right)^{-z}f
=\frac{1}{\Gamma\left(\frac{z}{2}\right)} \int_0^\infty
t^{\frac{z}{2}-1}h_t\star f\, dt.$$
Taking Fourier transforms we get 
$$\sigma_z=\frac{1}{\Gamma\left(\frac{z}{2}\right)} \int_0^\infty
t^{\frac{z}{2}-1} \widehat{h_t} dt$$ 
since $\widehat{h_t\star f}=\widehat{h_t}\cdot \hat f$. We therefore  need to compute the Fourier transform of $h_t$ and hence an explicit expression for the heat-kernel of the Laplace operator on $S^1$.   \\
 The heat kernel of the corresponding Laplace operator on $\R$ at time $t$ is given by $K_t(x, y)= k_t(x-y)$ with:
$$k_t(x):= \frac{1}{\sqrt{4\pi t}}e^{-\frac{x^2}{4t}}$$ and when identifying $S^1$ with $\R/ 2\pi \Z$, the heat-kernel of the Laplacian on  $S^1$ is given by 
$$H_t(x,y)=\sum_{n\in \Z} k_t(x-y+2\pi n).$$ 
The fact that it is "translation invariant  modulo $2\pi$" enables us to define the symbol using an ordinary Fourier transform. Setting $H_t(x, y)= h_t(x-y)$  we have: 
$$e^{-t\Delta} f= h_t\ast f\Rightarrow \widehat{e^{-t\Delta} f}= \hat h_t\, \hat f$$
so that the Fourier transform of $h_t$ can be intepreted as the symbol of $e^{-t\Delta}$. We first derive $h_t$ using the Poisson summation formula: 
$$\sum_{n\in\Z} f(x+n)= \sum_{k\in\Z} e^{2i\pi kx} \int_{-\infty}^{+\infty} f(y) e^{-2i\pi k y} dy.$$ Hence 
  \begin{eqnarray*}
h_t(x) &=&\sum_{n\in \Z} \tilde k_t(\frac{x}{2\pi}+ n)
\quad ( {\rm with} \quad \tilde k_t(y):=k_t(2\pi y))\\
&=& \sum_{k\in\Z} e^{i kx} \int_{-\infty}^{+\infty} k_t(2\pi y) e^{-2i\pi k y} dy\\
&=&\frac{1}{2\pi} \sum_{k\in\Z} e^{i kx} \int_{-\infty}^{+\infty} k_t( y) e^{-i k y} dy\\
&=& \frac{1}{2\pi\sqrt{4\pi t}} \sum_{k\in\Z} e^{i kx} \int_{-\infty}^{+\infty} e^{-\frac{y^2}{4t}} e^{-i k y} dy\\ 
&=& \frac{1}{2\pi} \sum_{k\in\Z} e^{i kx} e^{-{t k^2}}
\end{eqnarray*}
since for any $\lambda>0$ we have $ \int_{-\infty}^{+\infty}e^{-iy\xi} e^{-\frac{\lambda y^2}{2}}dy = \frac{\sqrt \pi}{\sqrt \lambda}e^{- \frac{1}{2\lambda}\xi^2}$. 
Considering any test function $\varphi\in C^\infty_c(\R)$ and taking Fourier
transforms we find: 
\begin{eqnarray*}
<\hat h_t,\,\varphi>&=&<h_t,\,\check \varphi>\\
&=&\int_{-\infty}^{+\infty}\sum_{k\in\Z}e^{-iky}e^{-tk^2}\check\varphi(y)\,dy\\
&=&\sum_{k\in\Z}e^{-tk^2}\int_{-\infty}^{+\infty}e^{-iky}\check
\varphi(y)\,dy\hbox{ (by Fubini's theorem)}\\
&=&\sum_{k\in\Z}\varphi(k)e^{-tk^2}.
\end{eqnarray*}
On the other hand the orthogonal projection $p$ on $\hbox{Ker }\Delta$
(i.e. the constant functions) is given
by:
$$p(f)(x)=\int_{S^1}f(y)\,dy.$$
Its kernel $K_p$ is then the constant function on $S^1\times S^1$ equal to
$1$. The associated function $\tilde K_p$ is the constant function $1$ on
$\R$, so the symbol of $p$ is the Dirac mass at $0$. From that we deduce that
the symbol of $e^{-t\Delta'}$ is given by:
$$\sum_{k\in\Z-\{0\}}e^{-tk^2}\delta_k.$$
Applying the Mellin transform we finally get:
\begin{eqnarray*}
\sigma_z(x,\,\xi)&=&\frac{1}{\Gamma(\frac z2)} \int_0^\infty t^{\frac z2-1}\sum_{k\in\Z-\{0\}}e^{-tk^2}\delta_k(\xi)\,dt\\
&=& \sum_{k\in\Z-\{0\}}|k|^{-z}\delta_k(\xi).
\end{eqnarray*}
\endsquare\\ \\
\subsection{Discrete sums of symbols  and the Euler-MacLaurin formula}
The symbol $\sigma_z$ just described involves Dirac measures so that we cannot
directly apply  the results of sections 2, 3 and 4 derived for smooth symbols to define its
truncated and regularised integrals. The presence of Dirac measures leads to
discrete sums which we need to truncate and regularise all the same; we therefore focus in this
paragraph on truncated and regularised discrete sums of symbols. \\ 
 As we shall see, the Euler-MacLaurin formula (\cite{Ha} Chap. 13) builds a
 bridge  between
 discrete sums on one hand and continuous integrals of symbols on the other
 hand. It enables to transpose the properties derived previously for
 regularised integrals and iterated nested integrals to regularised sums and
 iterated nested sums. 
Let us consider symbols $(x,\xi)\mapsto \sigma(x,\xi)$ of log-polyhomogeneous
symbols  on $\R$ in the class $CS^{*,k}$ (see section 1 and subsection 2.2)
 ``with constant coefficients'', i.e. independent of the first variable
 $x$. They clearly define  symbols on
the quotient $S^1=\R/2\pi\Z$ which we also call $\sigma$. We  drop the first variable $x\in S^1$ and consider $\sigma$
as a function of a single variable $\xi\in\R$ (here identified with $T^*_xS^1$
for any $x\in S^1$). Let us denote by  $CS^{*, k}(\R)$ the class of such
symbols and $CS^{*, *}(\R)$ the algebra generated by the union over $l\in \N$
of these sets. \\
There is a discretised version ${\cal P}$ of the Rota-Baxter  $P(\sigma)(\eta)=\int_{\vert \xi\vert \leq \vert \eta\vert }
\sigma(\xi)\, d\xi$  of section 4:
\begin{equation}\label{RB2}
{\cal P}(\sigma)(n)= \sum_{\vert k\vert\leq \vert n\vert} \sigma(k)\quad \forall
\sigma\in CS^{*,*}(\R),
\end{equation}
which has  properties similar to those of $P$ as the following lemma shows. 
\begin{lem}\label{lem:cutoff-discr}
For any $\sigma \in CS^{*, k}(\R)$, there is a symbol $\overline{{\cal
  P}(\sigma) } \in CS^{*, k+1}(\R)$ with same order  max$(0, \alpha+1)$
(where $\alpha$ is the order of $\sigma$) as $P(\sigma)$, which
interpolates  ${\cal P}(\sigma)$. More precisely, ${\cal P}(\sigma) (n)=\overline{{\cal
  P}(\sigma)}(n)\quad \forall n\in \N$  and  for
any $\sigma \in CS^{*, k}(\R)$, 
 $ P(\sigma)-\overline {{\cal P}(\sigma)}$ lies in $ CS^{*, k}(\R).$
\end{lem}
\begin{rk} Let $\sigma_1$ and $\sigma_2$ be two classical symbols of order
  $\alpha_1$, $\alpha_2$ respectively. It follows from the above lemma and
  Proposition 
\ref{prop:RotaBaxter}
 that
  $\sigma_1\, \overline{{\cal P}(\sigma_2)}$ has order $\alpha_1+{\rm
  max}(0, \alpha_2+1)$ so that if $\alpha_1<-1$ and $\alpha_2\leq -1$ it lies
in $L^1(\R)\cap CS^{*, 1}(\R)$.  
\end{rk}
{\bf Proof:}
  The results of subsection 2.1 and the Euler-MacLaurin formula are
the essential ingredients. We set $\tau(t):=\sigma(t)+\sigma(-t)$, so
that we have:
$${\cal  P}(\sigma)(m)=\sum_{k=0}^{m}\tau(k).$$
 Let us first recall the Euler-MacLaurin formula
(formula (13.6.3) in G.H. Hardy's monograph \cite{Ha}, with adapted  notations): Consider the Bernoulli numbers, defined by:
$$\frac {t}{e^t-1}=\sum_k\frac{B_k}{k!}t^k,$$
so that
$$B_0=1,\,B_1=-\frac 12,\,B_2=\frac 16,\,B_4=\frac 1{30},\ldots$$
and $B_{2k+1}=0$ for $k\ge 1$. Define for any $n$ the function $\phi_n$ by the equation:
\begin{equation}
t\frac {e^{xt}-1}{e^{t}-1}=\sum_{n\ge 1}\phi_n(x)\frac{t^n}{n!},
\end{equation}
and define $\psi_n$ as the $1$-periodic function equal to $\psi_n$ on the interval $[0,\,1[$. We then have for $N\in\N$:
\begin{eqnarray}\label{euler-maclaurin}
{\cal  P}(\sigma)(N)- P(\sigma)(N)&=&\sum_{m=0}^N\tau(m)-\int_0^N\tau(t)\,dt\nonumber\\
&=&\frac 12 \tau(N)
+\sum_{r=1}^k(-1)^{r-1}\frac {B_{2r}}{(2r)!}\tau^{(2r-1)}(N)
+C_k+T_{k,N}
\end{eqnarray}
with:
\begin{eqnarray}\label{constant}
C_k&=&\int_0^1\tau(t)\,dt+\frac 12 \tau(1)\nonumber\\
&-&\sum_{r=1}^k(-1)^{r-1}\frac
{B_{2r}}{(2r)!}\tau^{(2r-1)}(1)\nonumber\\
&-&\frac 1{(2k+2)!}\int_1^{+\infty}
\psi_{2k+2}(t)\tau^{(2k+2)}(t)\,dt
\end{eqnarray}
and:
$$T_{k,N}=\frac 1{(2k+2)!}\int_N^{+\infty}
\psi_{2k+2}(t)\tau^{(2k+2)}(t)\,dt.$$
Setting $$\overline{{\cal P}(\sigma)}(\xi):=P(\sigma)(\xi) +\frac 12 \tau( \xi)
+\sum_{r=1}^k(-1)^{r-1}\frac {B_{2r}}{(2r)!}\tau^{(2r-1)}( \xi)
+C_k+T_{k,{\vert \xi\vert}}$$
then  yields a symbol $\overline{{\cal P}(\sigma)}$ in $CS^{*, k+1}(\R)$. Indeed,
 we know by Proposition \ref{prop:RotaBaxter} in section 4 
that $P(\sigma)$ lies in $CS^{*, k+1}(\R)$ and has order max$(0, \alpha+1)$
where $\alpha$ is the order of $\sigma$. The other terms on the r.h.s
lie in $CS^{*, k}(\R)$ as a result of the fact that $\sigma$ itself lies in
$CS^{*, k}(\R)$ and have order $\leq \alpha$. Indeed,  since $\tau$ lies in $CS^{*,k}(\R)$, 
$\tau^{(2k+2)}$ also lies in $CS^{*,k}(\R)$ and   the remainder term
$\xi \mapsto T_{k,\vert \xi\vert}$ is arbitrarily smoothing.  \\
In particular, we see that $\overline{{\cal P}(\sigma)}-P(\sigma)$ lies in 
$CS^{*, k}(\R)$ and has order $\leq{\rm max}(0, \alpha)$ ($0$ is due to the
presence of the constant $C_k$) so that $\overline{{\cal
    P}(\sigma)}$ and $P(\sigma)$ have same order.\endsquare
\begin{rk}
Formula (\ref{euler-maclaurin})
applied for $k$ and $k+1$ respectively  shows $C_{k+1}=C_k$ so that
$C_k$ stabilises at a constant $C$ for large $k$.
\end{rk}
On the grounds of this result we set the following definition. 
\begin{defn}\label{cut-off-sum}
For any $\sigma \in CS^{*, k}(\R)$ the expression:
$$\cutoffsum_{k\in\Z}\sigma:=\hbox{\rm fp}_{N\to +\infty}\sum_{k=-N}^N
\sigma(k):=\hbox{\rm fp}_{R\to +\infty}\overline{{\cal P}(\sigma)}(R)$$
defines the {\rm cut-off sum\/} of $\sigma$ on the integers.
\end{defn}
\begin{rk} Since $\overline{{\cal P}(\sigma)}$ has same order as $P(\sigma)$,
  the sum $\sum_{k=-N}^N
\sigma(k)$  converges when the corresponding  integral $\int_{-N}^N
\sigma(\xi)\, d\xi$ converges, namely when
  $\sigma$ has order $<-1$ in which case we have: 
$$\cutoffsum_{k\in\Z}\sigma(k)=\sum_{k\in\Z}\sigma(k).$$
 
\end{rk}
 Let us now consider holomorphic perturbations of a symbol $\sigma \in CS^{*,
   k}(\R)$ (these are closely related to the ``gauged symbols'' of \cite{G2}). 
\begin{prop}\label{hol-discr} 
Let $z\mapsto \sigma_z$ be a holomorphic family of log-polyhomogeneous
symbols  on $\R$ of order $\alpha(z)=-qz+\alpha(0)$ with $q>0$ that lie in the class $CS^{*,k}(\R)$.
\begin{enumerate} 
\item
The cut-off sum:
\begin{equation}\label{fpsum}
\cutoffsum_{k\in\Z}\sigma_z(k):=
\hbox{\rm fp}_{R\to +\infty}\sum_{k=-R}^R \sigma_z(k)
\end{equation}
is a meromorphic function of $z$ which coincides with
$\sum_{k=-\infty}^{+\infty} \sigma_z(k)$ on the half-plane $\hbox{\rm Re
}z>\frac{\hbox{\rm Re }\alpha(0)+1}{q}$, with poles
in $\{\frac{\hbox{\rm Re }\alpha(0)+1-j}{q},\, q\in\N\}$ of order $\le l+1$.
\item
The difference:
$$\cutoffint_{\R} \sigma_z(\xi)\,d\xi-\cutoffsum_{k\in\Z}\sigma_z(k)$$
is a holomorphic function of $z$.
\end{enumerate}
\end{prop}
{\bf Proof:}
As can be seen from the expression of $\overline {{\cal P}(\sigma)}$, a
holomorphic perturbation of $\sigma$ in $CS^{*,k}(\R)$ induces a holomorphic
perturbation $\overline{ {\cal P}(\sigma)}(z):=\overline{ {\cal P}(\sigma_z)}$ of
$\overline {{\cal P}(\sigma)} $ in $CS^{*,k+1}(\R)$ which reads: 
$$ \overline{{\cal P}(\sigma)}(z)(\xi)=\int_{-\vert \xi \vert}^{\vert \xi\vert}\sigma_z(t)\,dt+\frac 12 \tau_z(\xi)
+\sum_{r=1}^k(-1)^{r-1}\frac {B_{2r}}{(2r)!}\tau_z^{(2r-1)}(\xi)
+C_k(z)+T_{k,\vert \xi \vert}(z)$$
where the various terms are obtained by substituting $\sigma_z$ to $\sigma$ in  the r.h.s. of
(\ref{euler-maclaurin}).
 By   Theorem \ref{thm:KV}
 the integral term $\int_{-\vert \xi
   \vert}^{\vert \xi\vert}\sigma_z$ shares all properties listed in
Proposition   \ref{hol-discr}. The term $\frac 12\tau_z(\xi)$ and each term
inside the sum yields a holomorphic family in 
the symbol class $CS^{*,k}(\R)$. The remainder term $\xi \mapsto
T_{k,\vert\xi\vert}(z)$ yields a holomorphic family of smoothing symbols.
Finally, formula (\ref{constant}) shows  that $C(z)$ is holomorphic in the half-plane:
$$H_k:=\{z\in\C,\,\hbox{\rm Re }z>\frac{1+\alpha(0)-2k-2}{q}\}.$$
As this holds for any $k$, the function $C(z)$ is holomorphic in the whole
complex plane, and Proposition \ref{hol-discr} is proven.
\endsquare\\ \\
As a fundamental example, consider the holomorphic family:
$$\sigma_z(\xi)=\chi(\xi)|\xi|^{-z},$$
where $\chi$ is a cut-off function which vanishes around $0$ and such that
$\chi(\xi)=1$ for $|\xi|\ge 1$. One gets the expected relation between the
cut-off sum of the symbol $\sigma_z$ and the zeta function:
\begin{cor}\label{cor:zetaS1}
We have the following equality of meromorphic functions with simple poles at integer numbers: 
$$ \cutoffsum_{k\in\Z} \sigma_z(k) \,   \\
= 2\, \zeta(z).$$
\end{cor}
{\bf Proof:} Since the cut-off sum coincides with the ordinary
sum of the series when it converges absolutely, the equality holds
for $z$ in the half-plane $\{\hbox{\rm Re }z>1\}$. By item 1. of Proposition
\ref{hol-discr} the cut-off sum is a
meromorphic function of $z$, which therefore coincides with the well-known
meromorphic continuation of $2\zeta$.\endsquare
\begin{rk}
A simple computation shows that the cut-off integral of $\sigma_z$ reads:
$$\cutoffint \sigma_z(\xi)\,d\xi=\frac{2}{z-1}+h(z),$$
where $h$ is holomorphic. We then recover from item 3. of Proposition
\ref{hol-discr} that $\zeta(z)-\frac{1}{z-1}$ is holomorphic in the whole
complex plane.
\end{rk}

\subsection{Discrete Chen sums of symbols}
Similarly to  the operator $P$, the  operator ${\cal P}$ satisfies relations
reminiscent of  Rota-Baxter relations of weight $-1$:$$
{\cal P}(\sigma)(n)\,{\cal P}(\tau)(n)={\cal P}( \sigma\,\overline{ {\cal P}(\tau)})(n)+
{\cal P}(\tau\, \overline{{\cal P}(\sigma)})(n)+ {\cal P}( \sigma \, \tau)(n)\quad
\forall n\in \N$$
with an extra  term ${\cal P}( \sigma \, \tau)$  that did not
arise in the weight zero Rota-Baxter relations for integrals we considered
previously. We want to  build from ${\cal P}$ discrete Chen sums of symbols inductively in a similar
manner to the way we built continuous Chen integrals of symbols from $P$. 
We first define from ${\cal P}$ the operators
\begin{eqnarray*}
{\cal P}_j: \hat\otimes_{i=1}^{j +1}CS(\R) &\to &  \hat\otimes_{i=1}^{j} {\rm  Map}(\N, \C)\\
 {\cal P}_j(\sigma) (n_1, \cdots,  n_{j})&:=& {\cal P}\left(\sigma (n_1, \cdots,  n_{j},\cdot)\right)(n_{j})\\
\end{eqnarray*} 
On the grounds of Lemma 
\ref{lem:cutoff-discr} we derive the following result. 
\begin{lem} \label{lem:iteratedsums}
Let $\sigma\in \hat \otimes_{i=1}^{j+1} CS(\R)$, then 
\begin{enumerate}
\item $\overline{{\cal P}(\sigma)}_j$ defined by 
$$
\overline{{\cal  P}(\sigma)}_j (\xi_1, \cdots,  \xi_{j}):=\overline{{\cal  P}\left(\sigma (\xi_1, \cdots,  \xi_{j},\cdot)\right)}(\xi_{j})$$
lies in $\hat\otimes_{i=1}^{j-1}CS(\R)\otimes CS^{*,1}(\R).$
\item Let $\sigma= \sigma_1\otimes \cdots \otimes \sigma_k \in \hat
  \otimes_{i=1}^{k} CS(\R)$, then $\overline{{\cal P}_1\circ\cdots \circ
    {\cal P}_{k-1}(\sigma_1\otimes \cdots \otimes \sigma_k)}$ defined
  inductively by 
 $$\overline{{\cal P}_1\circ\cdots \circ
    {\cal P}_{k-1}(\sigma_1\otimes \cdots \otimes \sigma_k)}:=
  \overline{{\cal P}\left(\overline{{\cal P}_2\circ\cdots \circ
{\cal P}_{k-1}(\sigma_1\otimes \cdots \otimes\sigma_{k}) }\right)}$$
lies in $ CS^{*,
    k-1}(\R)$ and has the same order as $ P_1\circ\cdots \circ
     P_{k-1}(\sigma_1\otimes \cdots \otimes \sigma_k)$, given by ${\rm max}(0,
     \cdots, {\rm max}(0, {\rm max}(0,
     \alpha_k+1)+\alpha_{k-1}+1),\cdots)+\alpha_2+n)+\alpha_1$ where $\alpha_i$
     is the order of $\sigma_i$. 
\end{enumerate}
\end{lem}
{\bf Proof:} The first assertion is a direct consequence of  Lemma 
\ref{lem:cutoff-discr}. The second assertion then follows from an induction
procedure on $j$ to check that 
$\overline{{\cal P}_{k-j}\circ\cdots \circ
    {\cal P}_{k-1}=(\sigma_1\otimes \cdots \otimes \sigma_k)}$ maps $\hat \otimes^k CS
   (\R)$ to $\hat \otimes^{k -j-1}CS
   (\R)\otimes CS^{*, j}(\R)$. The computation of the order also follows by
   induction using the fact that  by Lemma \ref{lem:cutoff-discr}, $\overline{ P(\sigma)}$ and $P(\sigma)$ have
   the same order derived in Theorem \ref{thm:iteratedRotaBaxter}.
\endsquare\\ \\
We are now ready to define discrete Chen sums of symbols. Combining Lemma
\ref{lem:iteratedsums}  with Lemma \ref{lem:cutoff-discr}
shows that the cut-off sum of the symbol $\overline{{\cal P}_1\circ\cdots \circ
    {\cal P}_{k-1}(\sigma_1\otimes \cdots \otimes \sigma_k)}$ is well defined
  so that we can set the following definition. 
\begin{defn} For $\sigma_1, \cdots, \sigma_k \in CS(\R)$, 
we call 
$$\cutoffsum^{Chen}\sigma_1\otimes \cdots \otimes \sigma_k:= \cutoffsum \overline{{\cal P}_1\circ\cdots \circ
    {\cal P}_{k-1}(\sigma_1\otimes \cdots \otimes \sigma_k)}$$
the cut-off Chen sum of $\sigma:=\sigma_1\otimes \cdots \otimes\sigma_k$.
\end{defn}
\begin{rk}
Given the expression of the order of $\overline{{\cal P}_1\circ\cdots \circ
    {\cal P}_{k-1}(\sigma_1\otimes \cdots \otimes \sigma_k)}$ explicited in
  the above lemma, it converges  whenever $\alpha_1<-1$ and $\alpha_i\leq -1$
  for all $i\neq 1$ in which case we have that
  $$\cutoffsum^{Chen}\sigma_1\otimes \cdots \otimes \sigma_k=
  \sum^{Chen}\sigma_1\otimes \cdots \otimes \sigma_k$$ is an ordinary discrete
  Chen sum. 
\end{rk}
\subsection{Multiple zeta functions}
We now apply the above results to $$\sigma_i:=\sigma_{s_i}:= \chi(\xi)\, |\xi|^{-s_i},$$
where $s_1, \cdots, s_k$ are real numbers and $\chi$ is a cut-off function which vanishes around $0$ and such that
$\chi(\xi)=1$ for $|\xi|\ge 1$. We want to generalise Corollary \ref{cor:zetaS1} to  integrals of tensor
products $\otimes_{i=1}^k \sigma_i(s_i)$ relating them to  multiple zeta functions
(investigated in \cite{H} and \cite{Z}, see also \cite{C} or \cite{Wa} for a
review on the subject). Applying  the results of the previous paragraph to the
$\sigma_i$'s of order $-s_i$ leads to the following result which gives back
a known domain of convergence for multiple zeta functions. 
\begin{thm}
If  $s_1>1$ and $s_i\geq 1$ for $i=2, \cdots, k$ the
discrete Chen sum  
 $\sum^{Chen}\sigma_{s_1}\otimes \cdots\otimes\sigma_{s_k}$
converges and is proportional to
the
multiple zeta function:$$ \sum^{Chen}\sigma_{s_1}\otimes
\cdots\otimes\sigma_{s_k}= 2^k\tilde \zeta(s_1,\cdots, s_k):= 2^k \sum_{1\leq n_{k}\leq n_{k-1}\leq \cdots\leq
  n_1} n_{k}^{-s_k} \cdots  n_1^{-s_1}.$$
It  extends to all $s_i\in \R$ by a cut-off Chen integral of
the type defined above: 
$$\tilde \zeta(s_1,\cdots, s_k):=2^{-k}\cutoffsum^{Chen}\sigma_{s_1}\otimes
\cdots\otimes\sigma_{s_k},$$ where we have used the same symbol for the
extended mutiple $\zeta$-function. 
 \end{thm}
{\bf Proof:} It follows immediately from applying the results of the
previous paragraph to $\sigma_i=\sigma_{s_i}$ of  order $-s_i$.\endsquare\\ \\  
As a consequence we can also write:
$$\tilde \zeta(s_1,\cdots, s_k)=\cutoffsum_{n=1}^\infty \, \tilde P_1\circ \cdots \circ
 \tilde P_{k-1}(\sigma_{s_1}\otimes \cdots \otimes \sigma_{s_k})(n)$$
where $$\tilde P(f)(m):=\sum_{1\leq n\leq m} f(m),\quad \forall f\in {\cal F}(\N,
\C)$$ and 
\begin{eqnarray*}
\tilde P_j: \hat\otimes_{i=1}^{j +1}{\rm  Map}(\N, \C) &\to &  \hat\otimes_{i=1}^{j} {\rm  Map}(\N, \C)\\
 \tilde  P_j(f) (n_1, \cdots,  n_{j})&:=& \tilde P\left(f (n_1, \cdots,  n_{j},\cdot)\right)(n_{j}).\\
\end{eqnarray*} 
If $s_1>1$ and $s_i\geq 1$ for $i\neq 1$ then clearly, we
have ordinary sums: 
$$\tilde \zeta(s_1,\cdots, s_k)=\sum_{n=1}^\infty \,\tilde  P_1\circ \cdots \circ
\tilde P_{k-1}(\sigma_{s_1}\otimes \cdots \otimes \sigma_{s_k})(n).$$
\begin{rk}
\begin{itemize}
\item 
One can check that the same type of results holds with the usual multiple zeta functions $$\zeta(s_1,\ldots,s_k):= \sum_{1\leq
   n-1<n_2<\cdots<n_k} n_k^{-s_k}\cdots n_1^{-s_1} $$
 instead of $\tilde\zeta(s_1,\ldots,s_k)$ provided the large inequalities
 between the $|\xi_j|$'s and $|n_j|$'s are replaced by strict ones.
\item The above results can be extended  \footnote{via an extra statement on Chen sums
  of holomorphic families which we omit here, but which can be established
  along the same lines as was the meromorphicity result on Chen integrals of
  holomorphic families.} to complex numbers $z_i$
  instead of real numbers $s_i$ replacing  $s_1\geq 1$ and
  $s_i>1, i\neq 1$ in   the convergence assumptions by  ${\rm Re}(z_1)\geq 1$ and
  ${\rm Re}(z_i)>1, i\neq 1$. 
\end{itemize}
\end{rk}
The  well known ``second shuffle
 relations'' for multiple zeta  functions \cite{ENR} come from the natural
 partition of  the domain:
$$P_{k,l}:=\{x_1>\cdots >x_k>0\} \times \{x_{k+1}>\cdots >x_{k+l}>0\}
\subset ]0,+\infty[^{k+l}$$
into:
$$P_{k,l}=\coprod_{\sigma\in\hbox{\eightrm mix sh}(k,l)}P_\sigma
,$$
where $\hbox{mix sh}(k,l)$ stands for the {\sl mixable shuffles\/}, i.e. the
surjective maps $\sigma$ from $\{1,\ldots k+l\}$ onto
$\{1,\ldots m(\sigma)\}$ (for some $m(\sigma)\le k+l$) such that $\sigma_1<\cdots <\sigma_k$ and
$\sigma_{k+1}<\cdots <\sigma_{k+l}$. The domain $P_\sigma$ is defined by:
$$P_\sigma=\{(x_1,\ldots ,x_{k+l})\,/\,
x_{\sigma_r}>x_{\sigma_{r+1}}\hbox{ if }\sigma_r\not =\sigma_{r+1}
\hbox{ and } x_r=x_{r+1} \hbox{ if } \sigma_r=\sigma_{r+1}\}.$$
The second shuffle relations are:
\begin{equation}\label{shuffle2}
\!\!\!\zeta_{k}(z_1,\cdots, z_k)\, \zeta_{l}(z_{k+1},\cdots ,z_{k+l})
=\!\!\!\sum_{\sigma\in \hbox{\eightrm mix
    sh}(k,l)}\zeta_{m(\sigma)}(Z_\sigma),
\end{equation}
where $Z_\sigma$ is the $m(\sigma)$-uple defined by:
$$Z(\sigma)_j=\sum_{i\in\{1,\ldots,k+l\},\, \sigma(i)=j}z_i.$$
For $k=l=1$ they read:
$$ \zeta(z_1) \zeta(z_2)=  \zeta(z_1, z_2)+ \zeta(z_2, z_1)+ \zeta(z_1+z_2).$$
Using the identification $ \cutoffint_{\R}\sigma_{z}(\xi)d\xi=2 \zeta(z)$
derived previously we can indeed compute: 
\begin{eqnarray*}
4 \zeta(z_1) \zeta(z_2)&=& \prod_{i=1}^2 \cutoffint_{\R}D(\sigma_{z_i})\\
&=&
\cutoffint_{\R}D(\sigma_{z_2})\int_{\vert\xi_1<\vert\xi_2\vert}D(\sigma_{z_1})+  \cutoffint_{\R}D(\sigma_{z_1})\int_{\vert\xi_2\vert <
  \vert\xi_1\vert}D(\sigma_{z_2}) \\
&+& \cutoffint_{\vert \xi_1\vert =\vert \xi_2\vert}D(\sigma_{z_1})\otimes D(\sigma_{z_2})\\
&=&  4 \zeta(z_1, z_2)+ 4\zeta(z_2, z_1)+ 4\zeta(z_1+z_2).  \\
\end{eqnarray*}
The verification of the general formula (\ref{shuffle2}) goes along the same lines.
\vfill \eject \noindent
\bibliographystyle{plain}

\end{document}